\documentclass[aps,prl,preprint,amsmath,amssymb,groupedaddress]{revtex4-1}

\usepackage{graphicx}
\usepackage{dcolumn}
\usepackage{bm}
\graphicspath{{figs/}}

\usepackage{color}

\newcommand{\comp}{\mathbf}
\newcommand{\dd}{\mathrm{d}}
\newcommand{\pp}{\mathrm{p}}
\newcommand{\Vc}{\comp{\hat{V}}}
\newcommand{\cc}{\mathrm{c}}
\newcommand{\uc}{\mathrm{uc}}
\newcommand{\cs}[1]{\comp{\hat{#1}}}
\newcommand{\m}{\mathrm{m}}
\newcommand{\ii}{\mathrm{i}}
\newcommand{\0}{\mathrm{in}}
\newcommand{\LL}{\mathrm{s}}

\newcommand{\s}{\mathrm{s}}

\renewcommand{\Re}{\mathrm{Re}}
\renewcommand{\Im}{\mathrm{Im}}

\newcommand{\Phic}{\comp{\hat{\Phi}}}
\newcommand{\pc}{\comp{\hat{p}}}

\graphicspath{ {../submitted/}{figs/} }

\definecolor{newtextcolor}{cmyk}{0.01,0.7,1,0} 

\definecolor{newtextcolor2}{cmyk}{1,0.01,1,0} 

\definecolor{newtextcolor3}{cmyk}{0.7,0.7,0.01,0} 

\begin{document}


\title{On $1/f^\alpha$ power laws originating from linear neuronal cable theory: power spectral densities of the soma potential, soma membrane current 
and single-neuron contribution to the EEG}
\tableofcontents


%
\author{Klas~H.~Pettersen \footnote{Corresponding author: klas.pettersen@umb.no}}
 \affiliation{Center for Integrative Genetics, Dept.~of Mathematical Sciences and Technology, Norwegian University of Life Sciences, {\AA}s, Norway.}
\author{Henrik Lind\'{e}n}%
\affiliation{Dept. of Computational Biology, School of Computer Science and Communication, Royal Institute of Technology (KTH), Stockholm, Sweden}
\affiliation{Dept.~of Mathematical Sciences and Technology, Norwegian University of Life Sciences, {\AA}s, Norway.}
\author{Tom Tetzlaff}
\affiliation{Inst. of Neuroscience and Medicine (INM-6), Computational and Systems Neuroscience Research Center, J\"ulich, Germany}
\affiliation{Dept.~of Mathematical Sciences and Technology, Norwegian University of Life Sciences, {\AA}s, Norway.}
\author{Gaute~T.~Einevoll}
\affiliation{Dept.~of Mathematical Sciences and Technology, Norwegian University of Life Sciences, {\AA}s, Norway}

\date{\today}
             

\begin{abstract}
Power laws, that is, power spectral densities (PSDs) exhibiting $1/f^\alpha$ behavior for large frequencies $f$, have commonly been observed in neural recordings. Power laws in noise spectra have not only been observed in microscopic recordings of neural membrane potentials and membrane currents, but also 
in macroscopic EEG (electroencephalographic) recordings. While complex network behavior has been suggested to be at the root of this phenomenon, we here demonstrate a possible origin of such power laws in the biophysical properties of single neurons described by the standard cable equation. Taking advantage of the analytical tractability of the so called ball and stick neuron model, we derive general expressions for the PSD transfer functions for a set of measures of neuronal activity: the soma membrane current, the current-dipole moment (corresponding to the single-neuron EEG contribution), and the soma membrane potential. These PSD transfer functions relate the PSDs of the respective measurements to the PSDs of the noisy input currents. With homogeneously distributed input currents across the
neuronal membrane we find that all PSD transfer functions express asymptotic high-frequency $1/f^\alpha$ power laws. The corresponding power-law exponents are analytically identified as $\alpha^{I}_\infty=1/2$ for the soma membrane current, $\alpha^{p}_\infty=3/2$ for the current-dipole moment, and $\alpha^{V}_\infty=2$ for the soma membrane potential. These power-law exponents are found for arbitrary combinations of uncorrelated and correlated noisy input current (as long as both the dendrites and the soma receive some uncorrelated input currents). Comparison with available data suggests that the apparent power laws observed in experiments may stem from uncorrelated current sources, presumably intrinsic ion channels, which are homogeneously distributed across the neural membranes and themselves exhibit pink ($1/f$) noise distributions.

The significance of this finding goes beyond neuroscience as it demonstrates how $1/f^\alpha$ power laws with a wide range of values for the power-law exponent $\alpha$ may arise from a simple, linear partial differential equation. We find here that the well-known cable equation describing the electrical properties of membranes transfers white-noise current input into 'colored' $1/f^\alpha$-noise where $\alpha$ may have any half-numbered value within the interval from $1/2$ to 3 for the different measurement modalities. Intuitively, the physical origin of these novel power laws can be understood in terms of the superposition of numerous low-pass filtered contributions with different cut-off frequencies (i.e., different time constants) due to the different spatial positions of the various current inputs along the neuron. As our model system is linear, the results directly generalize to any colored input noise, i.e., transferring  $1/f^\beta$ spectra of input currents to $1/f^{\beta+\alpha}$ output spectra.
\end{abstract}

\pacs{Valid PACS appear here}
\maketitle



\

\begin{Large}
\centerline{Popular summary}
\end{Large}

\

The common observation of power laws in nature and society, that is, that quantities or probabilities follow $1/x^\alpha$ distributions, has for long intrigued scientists. Such power laws have been seen in a wide range of situations including frequencies of differently sized earth quakes, distribution of links on the World Wide Web, and size scaling in animals. In the brain, power laws in the power spectral density (PSD) have been observed in electrophysiological recordings, both at the microscopic (single-neuron recordings) and macroscopic (EEG) levels. While these neural power laws have been suggested to stem from complex network behavior, we here demonstrate a possible origin of power laws in the basic biophysical properties of neurons, that is, in the standard cable-equation description of neuronal membranes. Taking advantage of the mathematical tractability of the so called ball and stick neuron model, we demonstrate analytically that high-frequency power laws in key experimental measures of neural activity will arise naturally when the noise sources are evenly distributed across the neuronal membrane. Comparison with available data further suggests that the apparent power laws observed in experiments may stem from uncorrelated current sources, presumably intrinsic ion channels, which are homogeneously distributed across the neural membranes and themselves exhibit pink ($1/f$) noise distributions. The significance of this finding goes beyond neuroscience as it demonstrates how $1/f^\alpha$ power laws with a wide range of values for the power-law exponent $\alpha$, i.e., any half-numbered value between $1/2$ and 3, may arise from a simple, linear physics equation.

\

\section{I. Introduction}

The apparent ubiquity of \emph{power laws} in nature and society, i.e., that quantities or probability distributions $y(x)$ satisfy the relationship
\begin{equation}
y(x) \propto x^{-\alpha}~,
\label{eq:power-law}
\end{equation}
where $\alpha$ is the power-law \emph{exponent}, has for a long time intrigued scientists~\cite{Stumpf2012}. Power laws in the tails of distributions have been reported in a wide range of situations including such different phenomena as frequency of differently sized earth quakes, distribution of links on the World Wide Web, paper publication rates in physics, and allometric scaling in animals (see \cite{Stumpf2012} and references therein). A key feature of power laws is that they are
\emph{scale invariant} over several orders of magnitude, i.e., that they do not give preference to a particular scale in space or time. There are several theories with such scale invariance as its fingerprint, among the most popular are fractal geometry \cite{Mandelbrot1977} and the theory of self-organized critical states \cite{Bak1987}.

Conspicuous power laws have been observed also in the field of neuroscience. Ever since Hans Berger recorded the first human electroencephalogram (EEG) in 1924 \cite{Berger1929}, its features have been under extensive study, especially since many of them are directly related to disease and to states of consciousness. Moreover, in the last decades the underlying \emph{power spectral density (PSD)} of the EEG has also attracted significant attention as the PSD is often well fitted by a $1/f^\alpha$ power law with $\alpha$ typically in the range from $1$ to $2.5$ \cite{Buzsaki2006,Freeman2003}.
Power-law spectra are not only seen in macroscopic neural recordings such as EEG, they also appear at the microscopic level, i.e., in single-neuron recordings. PSDs of the subthreshold membrane potentials recorded in the somas of neurons often resemble a $1/f^\alpha$ power law, typically with a larger exponent $\alpha$ ranging from 2 to 3 \cite{Diba2004,Rudolph2005,Jacobson2005, Yaron-Jakoubovitch2008,Bedard2008}. As for the EEG, this power law seems to be very robust: it has been observed across species, brain regions  and different experimental set-ups, such as cultured hippocampal layer~V neurons \cite{Diba2004}, pyramidal layer~IV--V neurons from rat neocortex \emph{in vitro}  \cite{Jacobson2005,Yaron-Jakoubovitch2008}, and neocortical neurons from cat visual cortex \emph{in vivo} \cite{Rudolph2005,Bedard2008}. At present, the origin, or origins, of these macroscopic and microscopic power laws observed in neural recordings are poorly understood.

Lack of sufficient statistical support have questioned the validity of identified power-law behaviors, and as a rule of thumb, a candidate power law should exhibit an approximately linear relationship in a log-log plot over at least two orders of magnitude~\cite{Stumpf2012}. Further, a mechanistic
explanation of how the power laws arise from the underlying dynamics should ideally be provided~\cite{Stumpf2012}. In the present paper we show through a combination of analytical and numerical investigations how power laws naturally can arise in neural systems from noise sources homogeneously distributed throughout neuronal membranes. We further show that the mechanism behind microscopic (soma potential, soma current) power laws will also lead to power laws in the single-neuron contribution (current-dipole moment) to the EEG, Moreover, if all single-neuron contributions to the recorded EEG signal exhibit the same power law, the EEG signal will also exhibit this power law. We find that for different measurement modalities different power-law exponents naturally follow from the well-established, biophysical cable properties of the neuronal membranes: the soma potential will be more low-pass filtered than the corresponding current-dipole moment determining the single-neuron contribution to the EEG \cite{Pettersen2008,Linden2010}, and as a consequence, the power-law exponent $\alpha$ will be larger for the soma potential than for the single-neuron contribution to the EEG~\cite{Hamalainen1993} (see illustration in Fig.~\ref{fig:fig1X}).

When comparing with experimental data, we further observed that for the special case when uncorrelated and homogeneously distributed membrane-current sources themselves exhibit $1/f$ power laws in their PSD, the theory predicts power-law exponents $\alpha$ in accordance with experimental observations for the \emph{microscopic} measures, i.e., the soma current and soma potential. The experimental situation is much less clear for the EEG signal. However, we note that under the assumption that such single-neuron sources dominate the high-frequency part of the EEG signal, the theoretical predictions are also compatible with the power-law-like behavior so far observed experimentally.

Both synaptic noise and intrinsic channel noise will in general contribute to the observed noise spectra, cf. Fig.~\ref{fig:fig1X}.
While our theory \emph{per se} is indifferent to the detailed membrane mechanism providing the noisy current, our findings suggests that the dominant noise source underlying the observed power spectra may be channel noise: prevalent theories for synaptic currents are difficult to reconcile with a $1/f$ power law, while potassium ion channels with such $1/f$ noise spectra indeed have been observed \cite{Derksen1966}.

Through the pioneering work by Wilfred Rall half a century ago \cite{Rall1959,Segev1994} the ball and stick neuron model was
established as a key model for the study of the signal processing properties of neurons. An important advantage is the model's analytical tractability, and this is exploited in the present study. We first demonstrate the relevance of this simplified model in the present context by numerical comparisons with results from a morphologically reconstructed multicompartmental pyramidal neuron model. Then we derive analytical power-law expressions for the various types of electrophysiological measurements.
While a single current input onto a dendrite does not give rise to power laws, we here show that power laws naturally arise for the case
with homogeneously distributed inputs across the dendrite and the soma~\cite{Tuckwell1983}, see Fig.~\ref{fig:fig1X}.
For this situation we show that the ball and stick neuron model acts as a power-law filter for high frequencies, i.e., the transfer function from the PSD of the input membrane currents, $s(f)$, to the PSD of the output (soma potential, soma current, or current-dipole moment setting up the EEG), $S(f)$, is described by a power law: $S(f)/s(f)=1/f^\alpha$. Notably the analytically derived power-law exponents $\alpha$ for these transfer functions are seen to be different for the different measurement modalities. The analytical expressions further reveal the dependence of the PSDs on single-neuron features such as the correlation of input currents, dendritic length and diameter, soma diameter and membrane impedance.

The theory presented here also contributes to $1/f$-theory in general: it illustrates that a basic physics equation, the cable equation, can act as a $1/f^\alpha$ power-law filter for high frequencies when the underlying model has spatially distributed input. Furthermore, $\alpha$ may have any half-numbered value between $1/2$ and $3$, depending on the physical measure (some potential, soma current, single-neuron contribution to the EEG) under consideration, and the coherence of the input currents. Intuitively, the emergence of the power-law spectra can be understood as a result of a superposition of low-pass filters with a wide range of cutoff frequencies due to position-dependent \emph{intrinsic dendritic filtering} \cite{Pettersen2008,Linden2010,Pettersen2012} of the spatially extended neuron.

The paper is organized as follows: In the next section we derive analytical expressions for the soma potential, soma current and current-dipole moment for the ball and stick neuron for the case with noisy current inputs impinging on the soma 'ball' and homogeneously on the dendritic stick. While these derivations are cumbersome, the final results are transparent: power laws are observed for all measurement modalities in the high-frequency limit. In Results we first demonstrate by means of numerical simulations the qualitative similarity of the power-law behaviors between the ball and stick model and a biophysically detailed pyramidal neuron. We then go on to analytically identify the set of power-law exponents for the various measurement modalities both in the case of uncorrelated and correlated current inputs. While the derived power laws strictly speaking refer to the functional form of PSDs in the high-frequency limit (Eq.~\ref{eq:power-law}), the purported power laws in neural data have typically been observed for frequencies less than a few hundred hertz. Our model study implies that the true high-frequency limit is not achieved at these frequencies. However, in our ball and stick model, quasi-linear relationships can still be observed in the characteristic PSD log-log plots for the experimentally relevant frequency range. These apparent power laws typically have smaller power-law exponents than their respective asymptotic value. The numerical values of these exponents will depend on details in the neuron model, but the ball and stick model has a very limited parameter space: it is fully specified by four parameters, a dimensionless frequency, the dimensionless stick length, the ratio between the soma and infinite-stick conductances, and the ratio between the somatic and dendritic current density. This allows for a comprehensive investigations of the apparent power-law exponents in terms of the neuron parameters, which we pursue next. To facilitate comparison with experiments we round off the Results section exploring how PSDs, and in particular apparent power laws, depend on relevant biophysical parameters. In the Discussion we then compare our model findings with experiments and speculate on the biophysical origin of the membrane currents underlying the observed PSD power laws.

\

\centerline{FIG. \ref{fig:fig1X} AROUND HERE}

\

\section{II. Models}
\label{sec:Neuron-models}
In the present study the idealized ball and stick neuron model will be treated analytically, while
simulation results will be presented for a reconstructed layer~V pyramidal neuron from cat visual cortex \cite{Mainen1996} (Fig.~\ref{fig:fig2X}).
Both the ball and stick model and the reconstructed layer V neuron model are purely passive, ensuring that linear theory can be used. The input currents are distributed throughout the neuron models with area density $\rho_\dd$ in the dendrite and $\rho_\s$ in the soma. The input currents share statistics, i.e., they all have the same PSD, denoted $s=s(\omega)$, and a pairwise coherence $c=c(\omega)$. The coherence is zero for uncorrelated input and unity for perfectly correlated input.

For the ball and stick neuron, the cable equation is treated analytically in frequency space. We first provide a solution for a single current input at an arbitrary position, and then use this solution as basis for the case of input currents evenly distributed throughout the neuronal membrane. The resulting PSDs can be expressed as Riemann sums where the terms correspond to single-input contributions. In the continuum limit where the neuron is assumed to be densely bombarded by input currents, the Riemann sums become analytically solvable integrals. From these analytical solutions we can then extract the various transfer functions relating the output PSDs to the PSDs of the input current. Here the output modalities of interest are the net somatic current, the soma potential and the single-neuron contribution to the EEG, see Figs.~\ref{fig:fig1X} and \ref{fig:fig2X}.

Below we treat the ball and stick neuron analytically. For the pyramidal neuron (Fig.~\ref{fig:fig2X}), the NEURON Simulation Environment \cite{Carnevale2006} with the supplied Python interface \cite{Hines2009} was used.

\

\centerline{FIG. \ref{fig:fig2X} AROUND HERE}

\

\subsection{A. Cable equation for dendritic sticks}
For a cylinder with a constant diameter $d$ the cable equation is given by
\begin{equation}
  \lambda^2 \frac{\partial^2 V(x,t)}{\partial x^2} = \tau_\m \frac{\partial V(x,t)}{\partial t} + V(x,t)~,
  \label{eq:cable}
\end{equation}
with the length constant $\lambda=1/\sqrt{g_\m r_\ii}=\sqrt{d R_\m/4R_\ii}$ and the time constant $\tau_\m=c_\m/g_\m=R_\m C_\m$.  $R_\m$, $C_\m$ and $R_\ii$ denote the specific membrane resistance, the specific membrane capacitance and the inner resistivity, respectively, and have dimensions $[R_\m]=\Omega \m^2$, $[C_\m]=\mathrm{F/m^2}$ and $[R_\ii]=\Omega \m$. Lower-case letters are used to describe the electrical properties per unit length of the cable: $g_\m=1/r_\m=\pi d/R_\m$, $c_\m=\pi d C_\m$ and $r_\ii=4R_\ii/\pi d^2$, with units $[g_\m]=1/\Omega\m$, $[c_\m]=\mathrm{F/m}$ and $[r_\ii]=\Omega/\m$. For convenience, the specific membrane conductance, $G_\m=1/R_\m$, will also be used, see Table~\ref{tab:glossary} for a list of symbols.

With dimensionless variables, $X=x/\lambda$ and $T=t/\tau_\m$, the cable equation, Eq.~\ref{eq:cable}, can be expressed
\begin{equation}
  \frac{\partial^2 V(X,T)}{\partial X^2} -\frac{\partial V(X,T)}{\partial T} - V(X,T) = 0~.
\end{equation}
Due to linearity, each frequency component of the input signal can be treated individually. For this, it is convenient to express the membrane potential in a complex (boldface notation) form,
\begin{equation}
  \comp{V} = \Vc(X,W) e^{j W T}~,
\end{equation}
where $\Vc$ is a complex number containing the amplitude $abs(\Vc)$ and phase $arg(\Vc)$ of the signal, and the dimensionless frequency is defined as $W=\omega \tau_\m$. The complex potentials are related to the measurable potential $V(X,T)$ through the Fourier components of the potential,
\begin{equation}
  V(X,T)=V_0(X)+\sum_{k=1}^\infty \Re \{ \Vc(X,W_k) e^{j W_k T} \}~,
\end{equation}
where $V_0(X)$ is the direct current (DC) potential.
The cable equation can then be simplified to
\begin{equation}
  \frac{\dd^2 \Vc}{\dd X^2}-\comp{q}^2 \Vc = 0~,
  \label{eq:cable_f}
\end{equation}
where $\comp{q}^2 \equiv 1+jW$, see \cite{Pettersen2008,Koch1998}. The general solution to Eq.~\ref{eq:cable_f} can be expressed as
\begin{equation}
  \cs{V}(X,W) = \comp{C}_1 \cosh (\comp{q}L-\comp{q}X)+\comp{C}_2 \sinh (\comp{q}L-\comp{q}X)~.
  \label{eq:gen_sol3}
\end{equation}
The expression for the axial current is given by
\begin{equation}
  I_\ii(x,t) = -\frac{1}{r_\ii} \frac{\partial V (x,t)}{\partial x}~,
\end{equation}
and is applied at the boundaries to find the specific solutions for the ball and stick neuron. In complex notation and with dimensionless variables this can be expressed as
\begin{equation}
  \cs{I}_\ii(X,W) = -\frac{1}{r_\ii \lambda} \frac{\partial \cs{V} (X,W)}{\partial X}=-G_\infty \frac{\partial \cs{V} (X,W)}{\partial X}~,
  \label{eq:axial_current}
\end{equation}
where $G_\infty$ is the infinite-stick conductance.
Similarly, the transmembrane current density
(including both leak currents and capacitive currents)
is given by
\begin{equation}
  i_\m =  -\frac{\partial I_\ii (x,t)}{\partial x} = \frac{1}{r_\ii}\frac{\partial^2 V(x,t)}{\partial x^2}~,
\end{equation}
with its complex counterpart,
\begin{equation}
  \cs{i}_\m(X,W) =- \frac{1}{\lambda} \frac{\partial \cs{I} (X,W)}{\partial X} =\frac{1}{r_\ii \lambda^2} \frac{\partial^2 \cs{V} (X,W)}{\partial X^2}=g_\m \frac{\partial^2 \cs{V} (X,W)}{\partial X^2}~.
  \label{eq:membrane_current}
\end{equation}

\

\centerline{FIG. \ref{fig:fig3X} AROUND HERE}

\

\subsection{B. Ball and stick neuron with single current input}
The ball and stick neuron \cite{Rall1959} consists of a dendritic stick attached to a single-compartment soma, see Fig.~\ref{fig:fig3X}\emph{A}. Here we envision the stick to be a long and thin cylinder with diameter $d$ and length $l$. The membrane area of the soma is set to be $\pi d_\s^2$, corresponding to the surface area of a sphere with diameter $d_\s$, or equivalently, the side area of a cylindrical box with diameter and height $d_\s$.

The solution of the cable equation for a ball and stick neuron with a single input current at an arbitrary dendritic position is found by solving the cable equation separately for the neural compartment proximal to the input current and the neural compartment distal to the input current,  These solutions are then connected through a common voltage boundary condition  $\cs{V}_0$ at the connection point. For the proximal part of the stick, Ohm's law in combination with the lumped soma admittance gives the boundary condition at the somatic site, and for the distal part of the stick, a sealed-end boundary is applied at the far end. In this configuration the boundary condition $\cs{V}_0$ acts as the driving force of the system. The potential $\cs{V}_0$ can, however, also be related to a corresponding input current $\cs{I}_\0$ through the input impedance,
i.e., $\cs{I}_\0=\cs{V}_0 \cs{Y}_\0$.

\subsubsection{Distal part of dendritic stick}
First, we focus on the part of the stick distally to the input in Fig.~\ref{fig:fig3X}\emph{A}. Assume that the stick has $\cs{V}_0$ as a boundary condition at the proximal end and a sealed-end boundary at the distal end. We use the subscript 'd' for \emph{distal} stick at the spatial coordinates, and shift the coordinate system so that the input is in $X_\dd=0$. The boundary condition at the proximal end, i.e., at the position of the input current, then becomes $\comp{V}(X_\dd=0)=\cs{V}_0$, while a sealed end is assumed at the distal end of the stick, i.e., at $X_\dd=L_\dd$. Here $L_\dd$ denotes the electrotonic length a the stick with physical length $l$, i.e., $L_\dd=l_\dd/\lambda$. A sealed-end boundary corresponds to zero axial current, Eq.~\ref{eq:axial_current}.

With these boundary conditions the specific solution to the cable equation becomes \cite{Koch1998,Pettersen2008},
\begin{equation}
  \cs{V}_\dd(X_\dd,W) = \frac{\cs{V}_0 \cosh (\comp{q}L_\dd-\comp{q}X_\dd)}{\cosh (\comp{q}L_\dd)}~.
  \label{eq:app_Xdpotential}
\end{equation}
The axial current $\cs{I}_\ii(X_\dd,W)$ is given by Eq.~\ref{eq:axial_current},
\begin{equation}
  \cs{I}_{\ii,\dd}(X_\dd,W) = \cs{V}_0 \comp{q} G_\infty  \frac{\sinh (\comp{q}L_\dd-\comp{q}X_\dd)}{\cosh (\comp{q}L_\dd)} ~.
\end{equation}
The dendritic input admittance, $\comp{Y}_{\mathrm{in},\dd}(W)=\cs{I}_{\ii,\dd}(X_\dd=0,W)/\cs{V}_\dd(X_\dd=0,W)$, will then be
\begin{equation}
  \comp{Y}_{\mathrm{in},\dd}(W) = \comp{q} G_\infty \tanh (\comp{q}L_\dd)~.
  \label{eq:impedance_input}
\end{equation}
Since $\mathop {\lim }\limits_{L \to \infty } \tanh (\comp{q}L)\rightarrow1$, the infinite-stick admittance can be expressed as $\comp{Y}_\mathrm{\infty}(W) = G_\infty \comp{q} = \comp{q}/r_\ii \lambda$, and the finite-stick admittance can be expressed as $\comp{Y}_{\mathrm{in},\dd}(W)=\comp{Y}_\mathrm{\infty}(W) \tanh (\comp{q}L_\dd)$.
From Eqs.~\ref{eq:membrane_current} and \ref{eq:app_Xdpotential} it follows that the transfer function linking an imposed voltage $\cs{V}_0$ in the proximal end to a transmembrane current density in position $X_\dd$ can be expressed as \cite{Pettersen2008}
\begin{equation}
  \cs{i}_{\m,\dd}(X_\dd,W)=g_\m \comp{q}^2 \frac{\cosh (\comp{q}L_\dd-\comp{q}X_\dd)}{\cosh (\comp{q}L_\dd)} \cs{V}_0~.
\end{equation}
%
The complex dipole-moment for a stick with a sealed end is then given by the integral
\begin{equation}
	\cs{p}_\dd(W)= \lambda^2 \int_0^{L_\dd} \cs{i}_{\m,\dd}(X,W) X~\dd X = \lambda G_\infty  \cs{V}_0 [1-1/\cosh (\comp{q}L_\dd)]~.
	\label{eq:dipole_moment_V}
\end{equation}
%

\subsubsection{Soma and proximal part of dendritic stick}
Let us now consider a ball and stick neuron with an input current at the far end of the stick, effectively corresponding to the proximal part of the ball and stick neuron in Fig.~\ref{fig:fig3X}\emph{A}. We denote the coordinates with the subscript 'p' for \emph{proximal}. Similar to the situation for the distal stick, we apply a boundary condition $\cs{V}_0$ to the site of the current input and put this in $X_\pp=0$, i.e., $\comp{V}_\pp(X_\pp=0)=\cs{V}_0$. The stick is assumed to lie along the $X_\pp$-axis, to have electrotonic length $L_\pp$, and the soma site located at $X_\pp=L_\pp$. The lumped-soma boundary condition implies that the leak current out of the dendritic end is, through Ohm's law, proportional to the soma admittance, $\cs{I}_{\ii,\pp}(L_\pp,W) =\cs{I}_\LL= \comp{Y}_\LL \cs{V}_\pp(L_\pp,W)=\comp{Y}_\LL \cs{V}_\s$, where $\cs{I}_\LL$, $\cs{V}_\s$ and $\comp{Y}_\LL$ denote the somatic transmembrane current, soma potential and somatic membrane admittance, respectively. Thus, for  $X_\pp=0$ the boundary condition becomes:
\begin{equation}
 \cs{V}_\pp(0,W)=\cs{V}_0~,
  \label{eq:Xeq0boundary}
\end{equation}
and, through Eq.~\ref{eq:axial_current}, we have at $ X_\pp=L_\pp$:
\begin{equation}
 \cs{I}_{\ii,\pp}(L_\pp,W)=\left.-G_\infty \frac{\partial \cs{V}_\pp (X_\pp,W)}{\partial X_\pp}\right|_{X_\pp=L_\pp}=\comp{Y}_\LL \cs{V}_\LL~.
  \label{eq:XeqLboundary}
\end{equation}
The complex constant $\comp{C}_2$ in Eq.~\ref{eq:gen_sol3} is found from the boundary condition in Eq.~\ref{eq:XeqLboundary},
\begin{equation}
  \comp{C}_2 = \frac{\comp{Y}_\LL \cs{V}_\LL}{G_\infty \comp{q}}= \cs{V}_\LL \frac{\comp{Y}_\LL}{\comp{Y}_\infty}~,
\end{equation}
which, combined with Eq.~\ref{eq:Xeq0boundary}, gives $\comp{C}_1$:
\begin{equation}
  \comp{C}_1 = \frac{\cs{V}_0}{\cosh (\comp{q} L_\pp)}-\cs{V}_\LL \frac{\comp{Y}_\LL}{\comp{Y}_\infty} \tanh (\comp{q}L_\pp)~.
\end{equation}
By substituting the constants $\comp{C}_1$ and $\comp{C}_2$ and by using $\cs{V}_\LL=\cs{V}(L_\pp,W)$, Eq.~\ref{eq:gen_sol3} gives
\begin{equation}
  \cs{V}_0/\cs{V}_\LL = \cosh (\comp{q}L_\pp)(1+\comp{Y} \tanh(\comp{q}L_\pp))~,
  \label{eq:pot_frac}
\end{equation}
where $\comp{Y}=\comp{Y}_\LL/\comp{Y}_\infty$.
Next, Eq.~\ref{eq:pot_frac} is used to substitute for $\cs{V}_\LL$ in the constants $\comp{C}_1$ and $\comp{C}_2$, and after some algebraic manipulations the solution for the cable equation with the given boundary conditions becomes,
\begin{equation}
  \cs{V}_\pp(X_\pp,W) = \cs{V}_0 \frac{\cosh(\comp{q}L_\pp-\comp{q}X_\pp)+\comp{Y} \sinh (\comp{q}L_\pp-\comp{q}X_\pp)}{\cosh(\comp{q}L_\pp)+\comp{Y} \sinh (\comp{q}L_\pp)}~.
  \label{eq:V}
\end{equation}
The axial current is through Eq.~\ref{eq:axial_current} given by
\begin{equation}
  \cs{I}_{\ii,\pp}(X_\pp,W) = \cs{V}_0 \comp{Y}_\infty \frac{\sinh (\comp{q}L_\pp-\comp{q}X_\pp)+\comp{Y} \cosh (\comp{q}L_\pp-\comp{q}X_\pp)}{\cosh (\comp{q}L_\pp)+\comp{Y} \sinh (\comp{q}L_\pp)}~,
\end{equation}
and the input admittance is, through Ohm's law, given by $\comp{Y}_{\mathrm{in},\pp}=\cs{I}_{\ii,\pp}(0,W)/\cs{V}_0$,
\begin{equation}
  \comp{Y}_{\mathrm{in},\pp} =  \comp{Y}_\infty \frac{\sinh (\comp{q}L_\pp)+\comp{Y} \cosh (\comp{q}L_\pp)}{\cosh (\comp{q}L_\pp)+\comp{Y} \sinh (\comp{q}L_\pp)}~.
\end{equation}
The axial current at $X_\pp=L_\pp$, i.e., the somatic transmembrane current, will then be
\begin{equation}
  \cs{I}_\LL=\cs{I}_{\ii,\pp} (L_\pp,W) = \frac{\cs{V}_0 \comp{Y}_\LL}
  {\cosh (\comp{q}L_\pp)+\comp{Y} \sinh (\comp{q} L_\pp)}~,
  \label{eq:Isoma}
\end{equation}
and the transmembrane current density will be given by Eq.~\ref{eq:membrane_current},
\begin{equation}
  \cs{i}_{\m,\pp} =  \cs{V}_0 g_\m \comp{q}^2 \frac{\cosh (\comp{q}L_\pp-\comp{q}X_\pp)+\comp{Y} \sinh (\comp{q}L_\pp-\comp{q}X_\pp)}{\cosh (\comp{q}L_\pp)+\comp{Y} \sinh (\comp{q}L_\pp)}~.
  \label{eq:im_lumped}
\end{equation}
By an integral similar to Eq.~\ref{eq:dipole_moment_V}, the current-dipole moment for the stick is found to be
\begin{equation}
	\cs{p}_\mathrm{stick}(W) = \cs{V}_0 \left[\lambda G_\infty-\frac{l_\pp \comp{Y}_\LL+\lambda G_\infty}{\cosh (\comp{q}L_\pp)+\comp{Y} \sinh (\comp{q}L_\pp)} \right]~.
  \label{eq:p_lumped}
\end{equation}
The contribution to the current-dipole moment from the somatic return current is the product of the somatic current, Eq.~\ref{eq:Isoma}, and the fixed dipole length (i.e., distance between the position of the current input and the soma), here corresponding  to the stick length $l_\pp$,
\begin{equation}
  \cs{p}_\LL=l_\pp \cs{I}_\LL = \frac{l_\pp \cs{V}_0 \comp{Y}_\LL} {\cosh (\comp{q}L_\pp)+\comp{Y} \sinh (\comp{q} L_\pp)}~.
\end{equation}
The total dipole moment for a ball and stick neuron with current input at the far end of the stick is therefore
\begin{equation}
  \cs{p}_\pp=\cs{p}_\LL+\cs{p}_\mathrm{stick} = \cs{V}_0\lambda G_\infty-\frac{\cs{V}_0 \lambda G_\infty} {\cosh (\comp{q}L_\pp)+\comp{Y} \sinh (\comp{q} L_\pp)}~.
  \label{eq:app_ppp}
\end{equation}

\subsubsection{Full solution}
The full solution for current inputs at arbitrary positions is achieved by superposition of the
distal-stick solution and the solution for the
proximal stick with a lumped soma, see Fig.~\ref{fig:fig3X}\emph{A}. We will now use the same notation and coordinate system as in Fig.~\ref{fig:fig3X}\emph{A}, i.e.,  $X_\pp=-X+L_\pp$ and $X_\dd=X-L_\pp$, and introduce the sum of the stick lengths $L=L_\pp+L_\dd$. Thus, the stick is along the $X$-axis from $X=0$ (soma end) to $X=L$ (distal end), and the input current is assumed to be injected at position $X'$. By summation of Eqs.~\ref{eq:dipole_moment_V} and \ref{eq:app_ppp} the ball and stick dipole moment now becomes
\begin{equation}
  \cs{p}=-\cs{V}_0 \lambda G_\infty \left[ \frac{1}{\cosh (\comp{q}L-\comp{q}X')}- \frac{1} {\cosh (\comp{q}X')+\comp{Y} \sinh (\comp{q} X')} \right]~.
  \label{eq:p_general_V0}
\end{equation}
The total input admittance of the ball and stick neuron is given by the sum of the proximal admittance and the distal admittance,
\begin{equation}
  \comp{Y}_\mathrm{in} = \comp{Y}_\mathrm{in,\pp}+\comp{Y}_{\mathrm{in},\dd}  = \comp{Y}_\infty \left[ \frac{\sinh (\comp{q}L_\pp)+\comp{Y} \cosh (\comp{q}L_\pp)}{\cosh (\comp{q}L_\pp)+\comp{Y} \sinh (\comp{q}L_\pp)}+  \tanh (\comp{q}L_\dd) \right]~,
  \label{eq:Yin_general1}
\end{equation}
which, with the coordinates used in Fig.~\ref{fig:fig3X}\emph{A}, becomes
\begin{equation}
  \comp{Y}_\mathrm{in} = \comp{Y}_\infty \left[ \frac{\sinh (\comp{q}X')+\comp{Y} \cosh (\comp{q}X')}{\cosh (\comp{q}X')+\comp{Y} \sinh (\comp{q}X')}+  \tanh (\comp{q}(L-X')) \right]~.
  \label{eq:Yin_general}
\end{equation}
From Eq.~\ref{eq:p_general_V0} we now find, by means of Ohm's law and this expression for the input admittance, the following transfer function between input current $\cs{I}_\0$ and dipole moment, $\cs{p}=\comp{T}_p \cs{I}_\0$,
\begin{equation}
	\comp{T}_p = \frac{\lambda G_\infty}{\comp{Y}_\infty} \frac{\cosh (\comp{q}L-\comp{q}X')-\comp{Y} \sinh (\comp{q}X')-\cosh (\comp{q}X')}{\comp{Y} \cosh(\comp{q}L)+\sinh (\comp{q}L)}~.
	\label{eq:appTp_single}
\end{equation}
Transfer functions for the other quantities of interest, $\comp{T}_V =\cs{V}_\s/\cs{I}_\0$, $\comp{T}_I =\cs{I}_\s/\cs{I}_\0$, $\comp{T}^\s_V =\cs{V}_\s/\cs{I}^\s_\0$, $\comp{T}^\s_I =\cs{I}_\s/\cs{I}^\s_\0$ , $\comp{T}^\s_p=\cs{p}_\s/\cs{I}^\s_\0$, can be found similarly.
The superscript 's' denotes that this applies for an input current at the soma. By substituting for $\cs{V}_0$ in Eq.~\ref{eq:Isoma}, the transfer function for the soma current becomes
\begin{equation}
	\comp{T}_I =\frac{\comp{Y}\cosh(\comp{q}L-\comp{q}X')}{\comp{Y} \cosh (\comp{q}L)+ \sinh (\comp{q}L)}~.
	\label{eq:appTI_single}
\end{equation}
From Eq.~\ref{eq:appTI_single} and by assuming Ohm's law for the soma membrane, the soma potential transfer function becomes
\begin{equation}
\comp{T}_V = \frac{1}{\comp{Y}_\infty} \frac{\cosh(\comp{q}L-\comp{q}X')}{\comp{Y} \cosh (\comp{q}L)+ \sinh (\comp{q}L)}~.
	\label{eq:appTV_single}
\end{equation}
For a somatic input current, $\cs{I}_\0=\cs{I}^\s_\0$, the soma potential is, through Ohm's law, described by its total neuron input impedance seen from soma,
\begin{equation}
		\comp{T}^\s_V= \frac{1}{\comp{Y}_\mathrm{in}(X'=0)}=\frac{1}{\comp{Y}_\infty}\frac{\cosh(\comp{q}L)}{\comp{Y}\cosh(\comp{q}L)+\sinh(\comp{q}L)}~.
	\label{eq:appTsV_single}
\end{equation}
By comparison between Eq.~\ref{eq:appTV_single} and Eq.~\ref{eq:appTsV_single}, we see that Eq.~\ref{eq:appTV_single} also applies for the special case with somatic input, i.e., $\comp{T}^\s_V=\comp{T}_V(X'=0)$. The \emph{net} somatic transmembrane current  (including both $\cs{I}_\0^\s$ and the somatic return current) has to enter the stick axially in $X=0$. Thus, the net somatic current can be described by $\cs{I}^\s_\s=-\cs{V}^\s_\s \left. \comp{Y}_{\mathrm{in},\dd} \right|_{L_\dd=L}$, and the transfer function becomes
\begin{equation}
	\comp{T}^\s_I=-\frac{\sinh(\comp{q}L)}{\comp{Y}\cosh(\comp{q}L)+\sinh(\comp{q}L)}~,
	\label{eq:appTsI_single}	
\end{equation}
which differs from the result in Eq.~\ref{eq:appTI_single}, i.e., $ \comp{T}^\s_I \ne \comp{T}_I (X'=0)$. The intracellular resistance between the soma and the start position $X=0$ of the stick is assumed to be zero, and the soma potential will therefore be the same regardless of whether the input current is positioned at the proximal end of the stick (i.e., at $X=0$) or in the soma. However, when estimating the net somatic membrane current this distinction is important: the current input will itself count as a part of the calculated soma current if it is positioned in the soma, but not if it is positioned at the proximal end of the dendritic stick.

For somatic input, the finite-stick expression in Eq.~\ref{eq:dipole_moment_V} will apply to the dipole moment. However, the input admittance is now different, and the transfer function becomes
\begin{equation}
 	\comp{T}^\s_p= \frac{\lambda G_\infty}{\comp{Y}_\infty} \frac{\cosh (\comp{q}L)-1}{\comp{Y} \cosh (\comp{q}L)+ \sinh (\comp{q}L)}~,
	\label{eq:appTsp_single}
\end{equation}
i.e., the expression in Eq.~\ref{eq:appTp_single} holds, $\comp{T}^\s_p=\comp{T}_p(X'=0)$.

\subsection{C. Ball and stick neuron with spatially distributed input}
\label{app:multi_syn_cable}
Above we derived transfer functions $\comp{T}$ for the ball and stick neuron, connecting current input at an arbitrary position on the neuron to the various measurement modalities, i.e., the current-dipole moment ($\comp{T}_p$), the soma potential ($\comp{T}_V$) and the soma current ($\comp{T}_I$). We will now derive expressions for the PSDs when the ball and stick neuron is bombarded with multiple inputs assuming that all input currents have the same PSD and a pairwise coherence $c(\omega)$ \cite{Linden2011}. The PSDs can then be divided into separate terms for uncorrelated ($c(\omega)=0$) and fully correlated ($c(\omega)=1$) input.

The PSD, $S=S(\omega)$, of the output can for the case of multiple current inputs be expressed as
\begin{eqnarray}
	 S&=&\sum_{k=1}^N\sum_{l=1}^N \cs{I}_\0^k\comp{T}^k (\cs{I}_\0^l \comp{T}^l)^*\nonumber\\
	 &=&s\left[(1-c)\sum_{k=1}^N \comp{T}^k (\comp{T}^k)^* + c \sum_{k=1}^N\sum_{l=1}^N \comp{T}^k (\comp{T}^l)^*\right]~\nonumber\\
	 &=& s\left[(1-c)\sum_{k=1}^N \left| \comp{T}^k \right|^2 + c \left| \sum_{k=1}^N \comp{T}^k \right|^2\right]=s H~,
	 \label{eq:PSDsums2}
\end{eqnarray}
where $s=s(\omega)$ is the PSD of the input currents, $c=c(\omega)$ is their coherence and $H=H(\omega)$ is the transfer function between the PSD of the input and the PSD of the output. The complex conjugate is denoted by the asterisk.

We now assume the first $J$ of the $N$ input currents to be positioned at the soma compartment, and the rest of the input to be spread homogeneously across the dendritic stick. The transfer function for the soma compartment, $\comp{T}^\s$, is the same for all somatic inputs, $\comp{T}^k=\comp{T}^\s$ for $k=1,2, \dots, J$, while the input transfer function for the dendritic stick is position dependent, $\comp{T}^k=\comp{T}(X_k,W)$ for $k=J+1,J+2, \dots, N$. The PSD transfer function can then be expressed
\begin{equation}
	 H=(1-c) \left(J\left| \comp{T}^\s\right|^2+\sum_{k=J+1}^N \left| \comp{T}^k\right|^2\right) + c \left| J \comp{T}^\s+\sum_{k=J+1}^N \comp{T}^k \right|^2~.
	 \label{eq:PSDsums3}
\end{equation}
To allow for analytical extraction of power laws, we next convert the sums into integrals. By assuming uniform current-input density (per membrane area) in the dendritic stick (given by $\rho_\dd=(N-J)/l \pi d$), it follows that the
axial density of current inputs is $1/(\rho_\dd \pi d)$. In the continuum limit ($N \rightarrow \infty$) we thus have
\begin{equation}
 \sum_{k=J+1}^N F(\comp{T}^k) \rightarrow \int_0^L F(\comp{T}(X)) \, \rho_\dd \, \pi d \; \lambda dX
 \label{eq:sum2integral}
\end{equation}
where the last factor $\lambda$ comes from the conversion to dimensionless lengths. The PSD transfer function, $H \equiv S/s$, in
Eq.~\ref{eq:PSDsums3} can then be split into three parts,
\begin{equation}
	H= \big(1-c\big) \big(H_{\uc,\s}+H_{\uc,\dd}\big)+ c \, H_\cc~,
	 \label{eq:PSDsummary}
\end{equation}
where
\begin{equation}
	H_{\uc,\s}=\rho_\s \pi d_\s^2 \left| \comp{T}^\s(W)\right|^2
		 \label{eq:appPSDucs}
\end{equation}
is the PSD transfer function for uncorrelated input at the soma compartment,
\begin{equation}
	H_{\uc,\dd}=\rho_\dd \pi d \lambda \int_0^L |\comp{T}(X,W)|^2 \dd X
		 \label{eq:appPSDucd}
\end{equation}
is the PSD transfer function for uncorrelated input distributed throughout the dendritic stick, and
\begin{equation}
	H_\cc=\left|\rho_\s \pi d_\s^2  \comp{T}^\s + \rho_\dd \pi d \lambda \int_0^L \comp{T}(X,W) \dd X \right|^2~
	\label{eq:appPSDc}
\end{equation}	
is the PSD transfer function for correlated input distributed both across the dendritic stick and onto the soma.

We have now derived (i) a general expressions for the PSD transfer function $H$ expressed by the general, single-input transfer functions $\comp{T}$ and $\comp{T}^\s$, and (ii) specific analytical expressions for the single-input transfer functions for the dipole moment, the soma potential and the soma current. We will next combine these results and analytically derive specific PSD transfer functions for the dipole moment, the soma potential and the soma current for distributed input.

\subsubsection{Correlated current inputs}

For correlated activity the somatic transfer function and the corresponding integral of the dendritic transfer function are summed, see Eq.~\ref{eq:appPSDc}. For the \emph{soma current} the integral within Eq.~\ref{eq:appPSDc} is given by
\begin{equation}
	\int_0^L \comp{T}_I(X,W) \dd X = \frac{\comp{Y} \sinh(\comp{q}L)/\comp{q}}{\comp{Y} \cosh(\comp{q}L)+\sinh(\comp{q}L)}~.
	\label{eq:int_I}
\end{equation}
By defining the denominator
\begin{equation}
	\comp{D}(\omega)=\comp{Y} \cosh (\comp{q}L)+ \sinh (\comp{q}L)~,
\label{eq:D}
\end{equation}
the PSD transfer function for the soma current is after some algebra found to be
\begin{eqnarray}
	H_\cc^I &=&  |(\rho_\dd \pi d \lambda \comp{Y}/\comp{q}-\rho_\s \pi d_\s^2) \sinh (\comp{q} L)|^2/|\comp{D}|^2 \nonumber\\
	&=& \frac{\pi^2 d_\s^4 (\rho_\dd -\rho_\s)^2}{2} [\cosh(2 a L)-\cos(2 b L)]/|\comp{D}|^2~,
	\label{eq:PSDIcRaw}
\end{eqnarray}
with the squared norm of $\comp{D}$ given by
\begin{eqnarray}
	 |\comp{D}|^2&=&\frac{1}{2} \left[ (B^2(a^2+b^2)+1) \cosh (2 a L)+2 a B \sinh (2 a L) \right. \nonumber\\
	 &+& \left. (B^2(a^2+b^2)-1 ) \cos (2 b L)+2 B b \sin (2 b L) \right]~,
	 \label{eq:PSDDn}
\end{eqnarray}
with $a$ and $b$ denoting the real and imaginary parts of $\comp{q}$, respectively, i.e.,
\begin{equation}
	a = ([(1+W^2)^{1/2}+1]/2)^{1/2}~,
	\label{eq:a}
\end{equation}
and
\begin{equation}	
	b = ([(1+W^2)^{1/2}-1]/2)^{1/2}~.
	\label{eq:b}
\end{equation}
In Eq.~\ref{eq:PSDDn} the specific membrane conductance and capacitance are assumed to be the same in the soma and the dendrite. Thus, $\comp{Y}_\LL= \pi d_\LL^2 \comp{q}^2 G_\m$ and $\comp{Y}_\mathrm{\infty} =\comp{q}/(\lambda r_\ii)$. The admittance ratio can then be expressed as
\begin{equation}
	\comp{Y}= \comp{q}B~,
\end{equation}
where $B=d_\s^2/(d \lambda)$.

The contribution to the \emph{soma potential} from dendritic input is given by the same integral as in Eq.~\ref{eq:int_I} divided by the somatic impedance.
By adding the corresponding transfer function for the somatic input the PSD transfer function is found to be:
\begin{eqnarray}
	H_\cc^V &=& |[\rho_\dd \pi d \lambda \sinh (\comp{q} L)/ \comp{q}+\rho_\s \pi d_\s^2 \cosh (\comp{q} L)]/\comp{Y}_\infty |^2/|\comp{D}|^2 \nonumber\\
	 &=&\frac{ \pi^2 \lambda^2 r_\ii^2}{2 \left(a^2+b^2\right)^2|\comp{D}|^2} \left[ \cos (2 b L) \left(d_\s^4 \rho_\s^2 \left(a^2+b^2\right)-d^2 \lambda^2 \rho_\dd^2\right) \right. \nonumber\\
	 &+& \cosh (2 a L) \left(d_\s^4 \rho_\s^2 \left(a^2+b^2\right)+d^2 \lambda^2 \rho_\dd^2\right) \nonumber\\
	 &+& \left. 2 d d_\s^2 \lambda \rho_\dd \rho_\s (a \sinh (2 a L)+b \sin (2 b L))\right]~.
	\label{eq:PSDVcRaw}
\end{eqnarray}

For the \emph{current-dipole moment}, the integral within Eq.~\ref{eq:appPSDc}, combined with the transfer function from Eq.~\ref{eq:appTp_single}, has the following simple solution,
\begin{equation}
	\int_0^L \comp{T}_p(X,W) \dd X=\frac{\lambda G_\infty}{\comp{Y}_\infty \comp{q} \comp{D}} \comp{Y}\left[1-\cosh(\comp{q}L) \right]~,
	\label{eq:int_p}
\end{equation}
and the PSD transfer function for the dipole moment for correlated input currents is found to be
\begin{eqnarray}
	H_\cc^p &=&  \left| \frac{\pi \lambda G_\infty [1-\cosh(\comp{q}L)] (\rho_\dd d \lambda \comp{Y}/\comp{q}-\rho_\s d_\s^2) }{\comp{Y}_\infty \comp{D}}   \right|^2 \nonumber \\
	 &=& \frac{\pi^2 d_\s^4 \lambda^2 (\rho_\dd-\rho_\s)^2 (\cos (b L)-\cosh (a L))^2}{(a^2+b^2) |\comp{D}|^2}~.
\end{eqnarray}

\subsubsection{Uncorrelated current inputs}

In the case of uncorrelated input currents, the squared norm of hyperbolic functions, as well as cross-terms of different hyperbolic functions, must be integrated from $X=0$ to $X=L$ to get the contributions from the dendritic stick.
These integrals can be solved by converting the hyperbolic functions to their corresponding exponential expressions and expanding the products before applying straight-forward integration of the different exponential terms. For example, the following integral has to be solved for all PSDs, both the soma current PSD, the soma potential PSD and the PSD of the single-neuron contribution to the EEG:
\begin{equation}
	I_1=\int_0^L |\cosh(\comp{q}L-\comp{q}X)|^2 \dd X~,
\end{equation}
where $I$ now denotes an integral, not a current. The integrand is translated to its exponential counterpart,
\begin{equation}
	I_1=\int_0^L \frac{1}{4} \left[ e^{(\comp{q}+\comp{q}^*)(L-X)}+e^{-(\comp{q}+\comp{q}^*)(L-X)}+ e^{(\comp{q}-\comp{q}^*)(L-X)}+ e^{-(\comp{q}-\comp{q}^*)(L-X)} \right] \dd X~,
\end{equation}
and the integral is straightforwardly evaluated and found to be:
\begin{align}
I_1=& \frac{1}{4}\left[ -\frac{1}{\comp{q}+\comp{q}^*}+\frac{1}{\comp{q}+\comp{q}^*}-\frac{1}{\comp{q}-\comp{q}^*}+\frac{1}{\comp{q}-\comp{q}^*} \right. \nonumber \\
& \left.  +\frac{e^{(\comp{q}+\comp{q}^*)L}}{\comp{q}+\comp{q}^*}-\frac{e^{-(\comp{q}+\comp{q}^*)L}}{\comp{q}+\comp{q}^*}+\frac{e^{(\comp{q}-\comp{q}^*)L}}{\comp{q}-\comp{q}^*}-\frac{e^{-(\comp{q}-\comp{q}^*)L}}{\comp{q}-\comp{q}^*}\right]~.
\end{align}
The expression can be transformed back to hyperbolic functions
\begin{equation}
	I_1=\frac{1}{2} \left( \frac{\sinh{[(\comp{q}+\comp{q}^*)L]}}{\comp{q}+\comp{q}^*}+\frac{\sinh{[(\comp{q}-\comp{q}^*)L]}}{\comp{q}-\comp{q}^*} \right)~,
\end{equation}
and simplified as
\begin{equation}
	I_1= \sinh (2aL)/4a+\sin (2bL)/4b~,
\end{equation}
where we have used
\begin{equation}
	\sinh (2 j bL) = j \sin (2bL)~.
\end{equation}

From the expressions for the single-input transfer functions for the \emph{soma potential}, Eq.~\ref{eq:appTV_single}, and \emph{soma current}, Eq.~\ref{eq:appTI_single}, it follows that $H_{\uc,\dd}^V$ and $H_{\uc,\dd}^I$ (cf.~Eq.~\ref{eq:appPSDucd}) are both proportional to $I_1$, i.e.,
\begin{equation}
	H_{\uc,\dd}^V=R_\infty^2 \frac{\sinh (2 a L)/a+\sin (2 b L)/b}{4 \left(a^2+b^2\right)}~,
\end{equation}
and
\begin{equation}
	H_{\uc,\dd}^I=\frac{B^2 (a^2+b^2) (a \sin(2 b L)+b \sinh(2 a L))}{4 a b}~.
\end{equation}
For $H_{\uc,\dd}^p$ the following integrals also appear:
\begin{eqnarray}
	I_2&=& \int_0^L |\cosh(\comp{q}X)|^2 \dd X~,\\
	I_3&=& \int_0^L |\sinh(\comp{q}X)|^2 \dd X~,\\
	I_4&=& \int_0^L \cosh(\comp{q}L-\comp{q}X) \cosh(\comp{q}^* X) \dd X~,\\
	\comp{I}_5&=& \int_0^L\cosh(\comp{q}L-\comp{q}X) \sinh(\comp{q}^* X) \dd X~,\\
	\comp{I}_6&=& \int_0^L \cosh(\comp{q}X) \sinh(\comp{q}^* X) \dd X~,
\end{eqnarray}
All integrals can be solved by a similar scheme as above, and the solutions are
\begin{eqnarray}
	I_2&=& \sinh{(2aL)}/4a+\sin{(2bL)}/4b~,\\
	I_3&=& \sinh(2aL)/4a- \sin{(2bL)}/4b~,\\
	I_4&=& \sinh(a L) \cos (b L)/2a+ \cosh(a L) \sin (b L)/2 b~,\label{eq:Isol4}\\
	\comp{I}_5&=&  \sinh(a L) \sin (b L)/2 b -j  \sinh(a L) \sin (b L)/2 a~,\label{eq:Isol5}\\
	\comp{I}_6&=&  \cosh (2 a L)/4a-1/4 a +j  \cos (2 b L)/4 b-1/4 b~.\label{eq:Isol6}
\end{eqnarray}
Note that the solutions to the integrals $\comp{I}_5$ and $\comp{I}_6$ are complex. In the expression for the dipole moment the complex conjugated versions of the integrals $\comp{I}_5$ and $\comp{I}_6$, i.e., $\comp{I}^*_5$ and $\comp{I}^*_6$, also appear. For these the results are found directly from Eqs.~\ref{eq:Isol5}-\ref{eq:Isol6} with  $j$ replaced by $-j$.
The PSD transfer function for the dipole moment with uncorrelated input at the dendrite only, $H_{\uc,\dd}^p$, can then be expressed as
\begin{equation}
	H_{\uc,\dd}^p=\frac{\rho_\dd  \pi d \lambda^3}{|\comp{q}|^2 |\comp{D}|^2} [I_1+I_2+|\comp{Y}|^2 I_3-2 \Re \{ \comp{I}_4 \}-2 \Re \{ \comp{Y^* \comp{I}_5}\}+2 \Re \{ \comp{Y}^* \comp{I}_6\}]~.
\end{equation}
The full expression of $H_{\uc,\dd}^p$ is then
\begin{eqnarray}
 	H_{\uc,\dd}^p &=&\frac{\rho_\dd  \pi d \lambda^3}{(a^2+b^2) |\comp{D}|^2} \left[\sinh{(2aL)}/2a+\sin{(2bL)}/2b \right. \nonumber\\
	&+&(y_1^2+y_2^2) (\sinh{(2aL)}/4a-\sin{(2bL)}/4b) \nonumber\\
	&-&  \sinh(aL) \cos(bL)/a+\cosh(aL)\sin(bL)/b  \nonumber\\
	&-& y_1 \sinh(aL) \sin(bL)/b+y_2 \sinh(aL) \sin(bL)/a \nonumber\\
	&+&\left.y_1 (\cosh (2aL) -1)/2a+y_2 (\cos (2bL) -1)/2b\right]~,\label{eq:appPSDpucd}
\end{eqnarray}
where $y_1= \Re \{ \comp{Y} \}$ and $y_2=\Im \{ \comp{Y} \}$.
For the special case where the specific admittance of the soma is equal to the specific admittance of the dendrite, i.e., $\comp{Y}=\comp{q}d_\s^2/\lambda d$, this simplifies to the expression given in Eq.~\ref{eq:PSDpucd}.

The somatic contributions to the uncorrelated PSD transfer functions are given by
\begin{equation}
	H_{\uc,\s}^I=\rho_\s \pi d_\s^2 [\cosh (2 a L)-\cos (2 b L)]/|\comp{D}|^2~,
\end{equation}
\begin{equation}
	H_{\uc,\s}^V=\frac{\rho_\s R_\m^2 d_\s^2}{\pi d^2 \lambda^2} \frac{\cosh(2 a L)+\cos(2 b L)}{2 (a^2+b^2)|\comp{D}|^2}~,
\end{equation}
and
\begin{equation}
	H_{\uc,\s}^p=\frac{\rho_\s \pi d_\s^2 \lambda^2}{2 (a^2+b^2)|\comp{D}|^2} [\cosh (2 a L)-2\cosh(a L) \cos(b L)+\cos (2 b L)+2]~,
\end{equation}
see Eqs.~\ref{eq:appTsV_single}-\ref{eq:appTsp_single}.

\subsection{D. Summary of PSD transfer functions for ball and stick neuron}
For convenience we here summarize the results, now solely in terms of dimensionless variables (except for the amplitudes $A$), i.e.,
$\rho \equiv \rho_\s/(\rho_\s+\rho_\dd)$, $B \equiv d_\s^2/(d\lambda)$, $L \equiv l/\lambda$, and $W \equiv \omega \tau$ (see Table~\ref{tab:series_parameters}).
The general expression for the PSD transfer functions reads:
\begin{equation}
	H = (1-c) H_{\uc} + c \, H_\cc~,	
	 \label{eq:PSDsummary-new}
\end{equation}
where $H_{\uc}=H_{\uc}(W)$ represents the contributions from uncorrelated current inputs,
$H_\cc=H_\cc(W)$ represents the contributions from correlated inputs, and $c=c(W)$ is the pairwise coherence function.
The contributions from uncorrelated input currents are in turn given as sums over contributions from somatic
$H_{\uc,\s}=H_{\uc,\s}(W)$ and dendritic inputs $H_{\uc,\dd}=H_{\uc,\dd}(W)$, i.e.,
\begin{equation}
	H_\uc = H_{\uc,\s} +  H_{\uc,\dd}~.	
	 \label{eq:PSDuncorrelated-new}
\end{equation}

The contribution to the PSD transfer functions for \emph{correlated} input currents are given by
\begin{eqnarray}
	H_\cc^I &=& A_\cc^I B^2  [\cosh(2 a L)-\cos(2 b L)]/|\comp{D}|^2~, \label{eq:PSDIc} \\
	H_\cc^p &=&  \frac{A_\cc^p B^2}{a^2+b^2} \left[ \cosh (2 a L)/2 \right. \nonumber\\
	& & -2 \left. \cosh(a L) \cos(b L)+\cos (2 b L)/2+1 \right]/|\comp{D}|^2~,\label{eq:PSDpc} \\
	H_\cc^V
	 &=& \frac{A_\cc^V}{2(a^2+b^2)^2} \left[ \cos (2 b L) \left(B^2 \rho^2 \left(a^2+b^2\right)-(1-\rho)^2\right) \right. \nonumber\\
	 & &+ \cosh (2 a L) \left(B^2 \rho^2 \left(a^2+b^2\right)+(1-\rho)^2\right) \nonumber\\
	 & &+ \left. 2 B (1-\rho) \rho (a \sinh (2 a L)+b \sin (2 b L))\right]/|\comp{D}|^2~,
	\label{eq:PSDVc}
\end{eqnarray}
with the squared norm of $\comp{D}$ given by Eq.~\ref{eq:PSDDn}, and $a$ and $b$ defined by Eqs.~\ref{eq:a} and \ref{eq:b}, respectively.

The contributions from \emph{uncorrelated} dendritic inputs are:
\begin{eqnarray}
	H_{\uc,\dd}^I &=&  \frac{A_{\uc,\dd}^I B^2 (a^2+b^2)}{\sqrt{2}} \left(\frac{\sinh{(2aL)}}{2a}+\frac{\sin{(2bL)}}{2b} \right)/|\comp{D}|^2~, \label{eq:uc_int_I} \\
 	H_{\uc,\dd}^p &=&\frac{A_{\uc,\dd}^p \sqrt{2}}{(a^2+b^2) } \left(\frac{\sinh{(2aL)}}{2a}+\frac{\sin{(2bL)}}{2b} \right. \nonumber\\
	& &+\frac{B^2(a^2+b^2)}{2} \left[\frac{\sinh{(2aL)}}{2a}-\frac{\sin{(2bL)}}{2b} \right] \nonumber\\
	& &-  \frac{\sinh(aL) \cos(bL)}{a}-\frac{\cosh(aL)\sin(bL)}{b}  \nonumber\\
	& &- \frac{B a \sinh(aL) \sin(bL) }{b}+\frac{B b \sinh(aL) \sin(bL) }{a} \nonumber\\
	& &+\left. B \frac{\cosh (2aL) -1}{2}+B \frac{\cos (2bL) -1}{2}\right)/|\comp{D}|^2~,
	\label{eq:PSDpucd} \\
	H_{\uc,\dd}^V &=& \frac{A_{\uc,\dd}^V B^2}{\sqrt{2}(a^2+b^2)} \left(\frac{\sinh{(2aL)}}{2a}+\frac{\sin{(2bL)}}{2b} \right)/|\comp{D}|^2~.
	\label{eq:uc_int_V}
\end{eqnarray}
In the special case with input to soma only, the PSD transfer functions are the same for uncorrelated (Eq.~\ref{eq:appPSDucs}) and correlated input (Eq.~\ref{eq:appPSDc}), the only difference being the amplitudes,
\begin{equation*}
	H_{\uc,\s}=\frac{\left. H_\cc \right|_{\rho=1}}{\rho_\s \pi d_\s^2}~.
	\label{eq:Ucs_Cs_relations}
\end{equation*}
($\rho= \rho_\s /(\rho_\s+\rho_\dd)=1$ implies that the input is onto soma only.)
The corresponding PSD transfer functions from uncorrelated somatic input thus become
\begin{eqnarray}
	H_{\uc,\s}^I &=& A_{\uc,\s}^I B^2  [\cosh(2 a L)-\cos(2 b L)]/|\comp{D}|^2~,
	\label{eq:PSDIucs} \\
	H_{\uc,\s}^p &=&  \frac{A_{\uc,\s}^p B^2}{a^2+b^2} \left[ \cosh (2 a L)/2 \right. \nonumber\\
	&-&2 \left. \cosh(a L) \cos(b L)+\cos (2 b L)/2+1 \right]/|\comp{D}|^2~,
	\label{eq:PSDpucs} \\
	H_{\uc,\s}^V &=& \frac{A_\cc^V B^2}{2 (a^2+b^2)} [\cosh (2 a L)+\cos (2 b L)] /|\comp{D}|^2~.
	\label{eq:PSDVucs}
\end{eqnarray}
%

\subsection{E. From single-neuron current-dipole moments to EEG}

In an infinite, homogenous, isotropic Ohmic medium with conductivity $\sigma$, the extracellular potential recorded at a given position $\vec{r}$ far away from a single-neuron current dipole is given by~\citep{Hamalainen1993,Nunez2006}.
\begin{equation}
    \Phi_1(\vec{r},t)= \frac{p_1(t) \cos \theta_1}{4 \pi \sigma (\vec{r}-\vec{r}_1)^2}~,
    \label{eq:singleneuron-Phi}
\end{equation}
where $\vec{r}_1$ designates the spatial position of the current dipole, $p_1$  is the magnitude of the current-dipole moment, and $\theta_1$ is the angle between the dipole moment vector $\vec{p}_1$ and the position vector $\vec{r}-\vec{r}_1$.
An important feature is that all time dependence of the single-neuron contribution to the potential $\Phi$ lies in $p_1(t)$ so that
$\Phi_1(\vec{r},t)$ factorizes as
\begin{equation}
    \Phi_1(\vec{r},t)= p_1(t) g_1(\vec{r})~.
    \label{eq:singleneuron-Phi_factor}
\end{equation}
For the electrical potential recorded at an EEG electrode, the forward model in Eq.~\ref{eq:singleneuron-Phi} is no longer applicable
due to different electrical conductivities of neural tissue, dura matter, scull and scalp. Analytical expressions analogous to Eq.~\ref{eq:singleneuron-Phi} can still be derived under certain circumstances such as with three-shell or four-shell concentric spherical head models (see~\citet{Nunez2006},~Appendix G), but the key observation for the present argument is that the single-neuron contribution to the EEG will still factorize, i.e.,  $\Phi_1(\vec{r},t)= p_1(t) \tilde{g}_1(\vec{r})$ where $\tilde{g}_1(\vec{r})$ here is an
unspecified function.

The compound EEG signal from a set of $N_n$ single-neuron current dipoles is now given by
\begin{equation}
    \Phi(\vec{r},t)=\sum_{n=1}^{N_n} p_n(t) \tilde{g}_n(\vec{r})~,
    \label{eq:EEGsum}
\end{equation}
where the index $n$ runs over all single-neuron current dipoles. For each Fourier component (frequency) we now have
\begin{equation}
  \Phic(\vec{r},f)=\sum_{n=1}^{N_n} \pc_n(f) \tilde{g}_n(\vec{r})~.
  \label{eq:EEGsum-Fourier}
\end{equation}
For the special case where the different single-neuron current dipoles moments are \emph{uncorrelated} we find
that the power spectral density $S^{EEG}_{UC}(f)$ of the EEG is of the form~\citep{Leski2013}
\begin{equation}
   S^{EEG}_{UC}(\vec{r},f) = |\Phic(\vec{r},f)|^2=\sum_{n=1}^{N_n} |\pc_n(f)|^2 \,|\tilde{g}_n(\vec{r})|^2~.
  \label{eq:PSD-EEG-uncorrelated}
\end{equation}
(We have here introduced the notation 'UC', i.e. capitalized, to highlight the difference between the present assumption of uncorrelated single-neuron current dipoles and the separate assumption of uncorrelated membrane currents onto individual neurons in the above sections.)
If the single-neuron current dipoles have the same power-law behavior in a particular frequency range, i.e.,
$|\pc_n(f)|^2 \approx c_n/f^{\alpha^p}$, it follows directly that the EEG signal will inherit this power-law behavior:
\begin{equation}
   S^{EEG}_{UC}(\vec{r},f) = \sum_{n=1}^{N_n} |\pc_n(f)|^2 \,|\tilde{g}_n(\vec{r})|^2 \approx
   \Big(\sum_{n=1}^{N_n} c_n \,|\tilde{g}_n(\vec{r})|^2 \Big)/f^{\alpha^p} = G_{UC}(\vec{r})/f^{\alpha^p} ~,
  \label{eq:PSD-EEG-uncorrelated-powerlaw}
\end{equation}
where $G_{UC}(\vec{r})$ determines the PSD amplitude, but not the slope.


The inheritance of the single-neuron power-law behavior also applies to
the case of correlated sources, provided that the pairwise coherences
are frequency independent. 
By similar reasoning as above we then find
\begin{equation}
  S^{EEG}_{C}(\vec{r},f) = \Big| \sum_{n=1}^{N_n} \pc_n(f) \,\tilde{g}_n(\vec{r}) \Big|^2 \approx
   G_{C}(\vec{r})/f^{\alpha^p} ~.
  \label{eq:PSD-EEG-correlated-powerlaw}
\end{equation}
Analogous expressions for the PSD for the EEG can also be derived when both correlated and uncorrelated single-neuron current dipoles
contribute, but we do not pursue this here; see \citet{Linden2011} and \citet{Leski2013} for more details.

\subsection{F. Numerical simulations}
The NEURON simulation environment \cite{Carnevale2006} with the supplied Python interface \cite{Hines2009} was used to simulate a layer-V
pyramidal neuron from cat visual cortex \cite{Mainen1996}. The main motivation for pursuing this was to allow for a
direct numerical comparison with results from the ball and stick neuron to probe similarities and differences, see Fig.~\ref{fig:fig2X}.
In addition, NEURON was also used on the ball and stick neuron model to verify consistency with the analytical results above.
Both the layer-V pyramidal neuron and the ball and stick neuron had a purely passive membrane, with specific membrane resistance $R_\m=3~\Omega \mathrm{m}^2$, specific axial resistivity $R_\ii=1.5~\Omega$m, and specific membrane capacitance $C_\m=0.01~$F/m$^2$. Simulations were performed with a time resolution of 0.0625~ms, and resulting data used for analysis had a time resolution of 0.25~ms. All simulations were run for a time period of 1200~ms and the first 200~ms were removed from the subsequent analysis to avoid transient upstart effects in the simulations.

The digital cell reconstruction of the layer-V pyramidal neuron was downloaded from ModelDB (\texttt{http://senselab.med.yale.edu/}), and the axon compartments were removed. To ensure sufficient numerical precision compartmentalization was done so that no dendritic compartment was larger than 1/30th of the electrotonic length at 100~Hz (using the function \texttt{lambda\_f(100)} in NEURON), which resulted in 3214 compartments. The soma was modeled as a single compartment.

The ball and stick neuron was modeled with a total of 201 segments, one segment was the iso-potential soma segment with length $20~\mathrm{\mu m}$ and diameter $20~\mathrm{\mu m}$, and 200 segments belonged to the attached dendritic stick of length 1~mm and diameter $2~\mathrm{\mu m}$.

Simulations were performed with the same white-noise current trace injected into each compartment separately.
The white-noise input current was constructed as a sum of sinusoidal currents \cite{Linden2010}
\begin{equation}
I(t) = I_0 \sum_{f=1}^{1000} \sin(2\pi ft+\varphi_f)
\end{equation}
where $\varphi_f$ represents a random phase for each frequency contribution.
Due to linearity of the cable equation, the contributions of individual current inputs could be combined to compute the PSD of the soma potential, the soma current and the dipole moment resulting from current injection into all $N$ compartments. In correspondence with Eq.~\ref{eq:PSDsums2}, the summation of the contributions from the input currents of different segments $i$  with membrane areas $A_i$ was done differently for uncorrelated and correlated input currents. The  uncorrelated PSDs, $S_\uc$, were computed according to
\begin{equation}
S_\uc(\omega) = \sum_{i=1}^N \rho_i A_i |\comp{y}_i(\omega)|^2~,
\end{equation}
while the correlated PSDs, $S_\cc$, were computed according to
\begin{equation}
S_\cc(\omega) = \left|\sum_{i=1}^N \rho_i A_i \comp{y}_i(\omega) \right|^2~.
\end{equation}
Here,  $\comp{y}_i(\omega)$ denotes the Fourier components of the signal $y(t)$ (either soma potential, soma current or dipole moment due to input in one segment), the product $\rho_i A_i$ gives the total number of input currents into one segment $i$, and the density $\rho_i$ represents $\rho_\dd$ for dendritic input and $\rho_\s$ for somatic input.

The total dipole moment $\vec{p}$ was in the numerical computations assumed to equal the dipole moment in one direction only: the direction along the stick for the ball and stick model, and the direction along the apical dendrite for the pyramidal neuron model, both denoted as the $x$-component, $p_x$. For the pyramidal neuron this is an approximation as the dipole moment also will have components in the lateral directions. However, the prominent 'open-field' asymmetry of the pyramidal neuron in the vertical direction suggests that this is a reasonable approximation when predicting contributions to the EEG signal.  The current-dipole moment is then given by
\begin{equation}
p_x= \sum_{i=1}^{N}x_i I_i(t)~,
\end{equation}
where $I_i$ is the transmembrane current of compartment~$i$, and $x_i$ is the corresponding $x$-position.

\section{III. Results}
\subsection{A. Biophysically detailed neuron model vs. ball and stick model}
To establish the relevance of using the simple ball and stick neuron to investigate the biophysical origin of power laws, we compare in Fig.~\ref{fig:fig2X} the normalized power spectral densities (PSDs) of the transmembrane soma current (row~1), the current-dipole moment (row~2), and the soma potential (row~3) of this model (column~1) with the corresponding results for a biophysically detailed layer-V pyramidal neuron (column~2); the rightmost column gives a direct comparison of PSDs. Both neuron models have a purely passive membrane and receive spatially distributed current input.
As described in the Models section (II.E), the PSD of the single-neuron contribution to the EEG will be proportional to the PSD of the neuronal current-dipole moment given the observation that the extracellular medium, dura matter, scull and scalp appear to be purely ohmic \cite{Nunez2006, Linden2010}. We here stick to the term 'current-dipole moment' even if the term 'single-neuron contribution to the EEG' could equally be used.

A first striking observation is that unlike single-input PSDs (thin gray lines in Fig.~\ref{fig:fig2X}), the PSDs resulting from numerous, homogeneously distributed input currents (thick lines) have a linear or quasi-linear appearance for high frequencies in these log-log plots, resembling $1/f^\alpha$ power laws. This is seen both when the numerous current inputs are correlated
(green thick lines) and uncorrelated (blue thick lines). We also observe that the decay in the PSD with increasing frequency is strongest for the soma potential, somewhat smaller for the current-dipole moment, and smallest for the soma current. This is reflected in the power-law exponents $\alpha$ estimated at 1000~Hz from these PSDs, see legend in Fig.~\ref{fig:fig2X}.  Here we observe that $\alpha$ is largest for the soma potential (bottom row) and smallest for the soma current (top row).

In the example in Fig.~\ref{fig:fig2X} we have assumed constant input current densities across the neurons, i.e., $\rho_\s=\rho_\dd$.
For this special case, correlated current input will, at all times, change the membrane charge density equally across the neuron, and as a consequence the neuron will be iso-potential. In this case the axial current within the neuron will be zero, and likewise the net membrane current (with the capacitive current included) for any compartment, including the soma. As a consequence the current-dipole moment vanishes, and the model can effectively be collapsed to an equivalent single-compartment neuron. For the soma current and dipole moment we thus only show results for uncorrelated inputs in Fig.~\ref{fig:fig2X}. However, correlated current input will still drive the soma potential (green curves in columns 1 and 2). Here we observe that the exponent $\alpha$ is smaller for uncorrelated input than for correlated input both for the ball and stick neuron and for the pyramidal neuron.

The results above pertains to the situation with white-noise current inputs, i.e., flat-band PSDs.
However, the results are easily generalized to the case with current inputs with other PSDs. Since our neuron models are passive and thus linear, the PSDs simply multiply. This is illustrated in column~3 of Fig.~\ref{fig:fig2X} which shows how our PSDs for uncorrelated input change with varying PSDs of the current input, $s(\omega)$. The blue curves correspond to white-noise input and are identical to the blue curves in column~2. The pink and brown curves illustrate the case of pink ($1/f$) and Brownian ($1/f^2$) input, respectively. Since the PSDs multiply, the power-law exponent of the input noise simply adds to the exponent $\alpha$. Thus, the pink and Brownian input increase the slope $\alpha$ with $1$ and $2$, respectively, compared to white-noise input.

Even though the dendritic structure of the reconstructed pyramidal neuron is very different from the ball and stick neuron in that it has both a highly branched structure and a varying diameter along its neural sections (tapering), both models seem to produce linear or quasi-linear high-frequency PSDs in the log-log representation. Also the power-law exponents are found to be fairly similar. This implies that the ball and stick neuron model captures salient power-law properties of the more biophysically detailed neuron model, and motivates our detailed analytical investigation of the power-law properties of the ball and stick neuron following next.

\subsection{B. Power laws for ball and stick neuron}

In the Models section above we derived analytical expressions for the PSD transfer functions of the soma current ($H^{I}$), current-dipole moment ($H^{p}$) and soma potential ($H^{V}$) for the ball and stick neuron for spatially distributed input currents. The resulting transfer functions $H(f)$, summarized in Eqs.~\ref{eq:PSDsummary-new}-\ref{eq:PSDVucs}, were of the form
\begin{equation}
	H(f)= \big(1-c(f)\big) \big(H_{\uc,\s}(f) + H_{\uc,\dd}(f)\big) + c(f) \, H_\cc(f)~,	
	 \label{eq:PSDsummary2}
\end{equation}
where $H_{\uc,\s}(f)$ and $H_{\uc,\dd}(f)$ represent the contributions from uncorrelated somatic and dendritic inputs, respectively, and $H_{\cc}(f)$ represents the contribution from correlated inputs.
$c=c(f)$ is the pairwise coherence of the current inputs, all assumed to have the same PSDs ($s=s(f)$).

These mathematical expressions are quite cumbersome, but they are dramatically simplified in the high-frequency limit, $f \rightarrow \infty$, in which the dominant power can be found analytically by a series expansions of the mathematical expressions for the transfer functions in Eqs.~\ref{eq:PSDIc}-\ref{eq:PSDVucs}.


The expressions for the PSD transfer functions contain terms which are both polynomial and superpolynomial (i.e., including
exponentials/exponentially decaying functions) with respect to frequency. As these superpolynomial terms will dominate the polynomial terms in the high-frequency limit, it follows from Eq.~\ref{eq:PSDDn} that for high frequencies the absolute square of the denominator $\comp{D}$ can be approximated by
\begin{equation}
	|\comp{D}|^2 \approx \sinh(2 aL) \left[\coth (2 a L)(B^2(a^2+b^2)+1)/2+a B \right]~,
\end{equation}
where terms decaying exponentially to zero with increasing frequency have been set to zero. The frequency dependence is through $a$ and $b$, see Eqs.~\ref{eq:a} and \ref{eq:b}. Note that $\lim_{f \rightarrow \infty} \coth (2 a L) = 1$ since $\lim_{f \rightarrow \infty} a = \infty$. In the high-frequency limit the PSD transfer functions Eqs.~\ref{eq:PSDIc}-\ref{eq:PSDVucs} become
\begin{eqnarray}
	H_\cc^I &\approx& A_\cc^I/(a^2+b^2+2 a/B+1/B^2)~,\label{eq:appPSDIc_superpoly}\\
	H_\cc^p &\approx& A_\cc^p/[(a^2+b^2)(a^2+b^2+2 a/B+1/B^2)]~,\\
	H_\cc^V &\approx& A_\cc^V \frac{\rho^2 [B^2(a^2+b^2)+1-2 a B]+2 \rho (a B-1)+1}{(a^2+b^2)^2[B^2(a^2+b^2)+2 a B+1]}~,\label{eq:appPSDVc_superpoly}\\
	H_{\uc,\dd}^I &\approx& A_{\uc,\dd}^I (a^2+b^2)/[ \sqrt{2} a (a^2+b^2+2 a/B+1/B^2)]~,\\
    H_{\uc,\dd}^p &\approx& A_{\uc,\dd}^p\frac{a^2+b^2-2 a/B+2/B^2}{\sqrt{2} a (a^2+b^2)(a^2+b^2+2 a/B+1/B^2)}~,
    \label{eq:appPSDVuc_superpoly} \\	
    H_{\uc,\dd}^V &\approx& A_{\uc,\dd}^V/[ \sqrt{2} a (a^2+b^2)(a^2+b^2+2 a/B+1/B^2)]~,
\end{eqnarray}
where the amplitudes $A$ are found in Table~\ref{tab:series_parameters}. When the PSDs expressed in Eqs.~\ref{eq:appPSDIc_superpoly}-\ref{eq:appPSDVuc_superpoly} are expanded reciprocally for high frequencies, i.e.,
$W =\omega \tau_\m = 2 \pi f \tau_\m \gg 1$, we get
\begin{eqnarray}
	H_{\uc,\dd}^I/A_{\uc,\dd}^I &\approx& 1/[W^{1/2}+\sqrt{2}/B+(1/B^2+1/2) W^{-1/2}+\mathcal{O}(W^{-1})]~,\label{eq:PSDIuc_asymptotic}\\
	H_{\cc}^I/A_{\cc}^I &\approx& 1/[W+\sqrt{2}W^{1/2}/B+1/B^2+\mathcal{O}(W^{-1/2})]~,\label{eq:PSDIc_asymptotic}\\
	H_{\uc,\dd}^p/A_{\uc,\dd}^p &\approx& 1/[(W^{3/2}+2 \sqrt{2} W/B+(B^2+6) W^{1/2}/2 B^2+\mathcal{O}(W^{0})]~,\label{eq:PSDPuc_asymptotic}\\
	H_{\cc}^p/A_{\cc}^p &\approx& 1/[W^2+\sqrt{2}W^{3/2}/B+W/B^2+\mathcal{O}(W^{1/2})]~,\label{eq:PSDPc_asymptotic}\\
	H_{\uc,\dd}^V/A_{\uc,\dd}^V &\approx& 1/[W^{5/2}+W^2  \sqrt{2}/B+W^{3/2}(1/B^2+1/2) +\mathcal{O}(W^{1})]~\label{eq:PSDVuc_asymptotic},\\	 H_{\cc}^V/A_{\cc}^V &\approx& 1/[W^2/\rho^2+W^{3/2}\sqrt{2}(2 \rho -1)/B \rho^3+ W (1-2\rho)^2/B^2 \rho^4+\mathcal{O}(W^{1/2})] \label{eq:PSDVc_asymptotic}~,
\end{eqnarray}
where $\rho$ is the dimensionless relative density, $\rho=\rho_\s/(\rho_\s+\rho_\dd)$, and $B=d^2_\s/\lambda d$, with $d_\s$ and $d$ denoting the somatic and dendritic diameter, respectively, and $\lambda$ denoting the dendritic length constant. The expansions were done in Mathematica (version 7.0), and a list of parameters used throughout the present paper is given in Table~\ref{tab:glossary} (along with the default numerical values used in the numerical investigations in later Results sections).

In Eqs.~\ref{eq:PSDIuc_asymptotic}-\ref{eq:PSDVc_asymptotic} terms which are exponentially decaying to zero for large $W$ have been approximated to zero. Note that Eq.~\ref{eq:PSDVc_asymptotic} does not apply in the special case of no somatic input, $\rho=0$, for which the series expansion gives
\begin{equation}
	H_\cc^V/A_\cc^V \approx 1/[W^3 B^2+W^{5/2} \sqrt{2} B+W^2+\mathcal{O}(W^{3/2})]~.\label{eq:PSDVc_asymptotic_rho0}
\end{equation}
The corresponding high frequency expansions of the PSD transfer functions for uncorrelated somatic input, $H_{\uc,\s}/A_{\uc,\s}$, are not shown, as these expressions are identical to the corresponding transfer functions for correlated input into the soma only, $H_\cc/A_\cc$ (i.e., equal to Eqs.~\ref{eq:PSDIc_asymptotic}, \ref{eq:PSDPc_asymptotic} and \ref{eq:PSDVc_asymptotic} with
$\rho=1$).

Eqs.~\ref{eq:PSDIuc_asymptotic}-\ref{eq:PSDVc_asymptotic_rho0} show that, due to position-dependent frequency filtering of the numerous inputs spread across the membrane (cf. Fig.~\ref{fig:fig3X}B), all PSD transfer functions express asymptotic high-frequency power laws. Moreover, these genuine 'infinite-frequency' power-law exponents, denoted $\alpha_\infty$, span \emph{every half power} from $\alpha_\infty=1/2$ (for $H_{\uc,\dd}^I$, Eq.~\ref{eq:PSDIuc_asymptotic}) to $\alpha_\infty=3$ (for $H_\cc^V$, Eq.~\ref{eq:PSDVc_asymptotic_rho0}) for the different transfer functions. The results are summarized in Table~\ref{tab:series_parameters}.

To obtain the power-law exponents in the general case with contributions from both correlated and uncorrelated current inputs, we need to
compare the different terms in the the general expression for $H(f)$ in Eq.~\ref{eq:PSDsummary2}. With different leading power-law exponents
$\alpha_\infty$ in their asymptotic expressions, the term with the lowest exponent will always dominate for sufficiently high frequencies.
From Table~\ref{tab:series_parameters} we see that for all three quantities of interest, i.e., $H^I(f)$, $H^p(f)$ and $H^V(f)$, the lowest exponent always comes from contributions from uncorrelated inputs. Note that the correlated term in Eq.~\ref{eq:PSDsummary2} also involves a frequency-dependent coherence term $c(f)$, but to the extent it modifies the PSD, it will likely add an additional low-pass filtering effect
\cite{Leski2013} and, if anything, increase the power-law exponent.
If we assume that the coherence is constant with respect to frequency we identify the following asymptotic exponents $\alpha^\mathrm{all}_\infty$ (i.e., with 'all' types of possible input) for $H^I$, $H^p$ and $H^V$:
\begin{equation*}
\alpha^{\mathrm{all},I}_\infty=1/2, \; \alpha^{\mathrm{all},p}_\infty=3/2, \; \alpha^{\mathrm{all},V}_\infty=2 \; .
\end{equation*}
Note that these power-law exponents are unchanged as long as uncorrelated activity is distributed both onto the soma and the dendrite, but will increase to $\alpha^I_\infty=1$ and $\alpha^p_\infty=2$ if no uncorrelated input are present on the dendrite. Similarly, without input onto soma, the asymptotic value will change for the soma potential PSD: it becomes $\alpha^V_\infty=2.5$ if uncorrelated input is uniformly
distributed on the dendrite, and $\alpha^V_\infty=3$ if the dendritic input is correlated.

\subsection{C. Apparent power laws for experimentally relevant frequencies}

Detailed inspection of the power-law slopes for the ball and stick model in Fig.~\ref{fig:fig2X} and comparison with the power-law exponents $\alpha_\infty$ listed in Table~\ref{tab:series_parameters} reveal that although the curves might look linearly decaying
in the log-log plot for high frequencies, the expressed exponents $\alpha$ are still deviating from their high-frequency values $\alpha_\infty$, even at 1000~Hz. As experimental power laws have been measured for much lower frequencies than this, we now go on to investigate \emph{apparent} PSD power laws for lower frequencies. For this it is convenient to define a low-frequency (\emph{lf}) regime, an intermediate-frequency (\emph{if}) regime and a high-frequency (\emph{hf}) regime, as illustrated in Fig.~\ref{fig:fig3X}\emph{C}. The transition frequencies between the regimes are given by the frequencies at which $\alpha$ is $50\%$ and $90\%$ of $\alpha^\mathrm{all}_\infty$, respectively.

The log-log decay rates of the PSD transfer functions can be defined for any frequency by defining the slope $\alpha(W)$ as the negative log-log derivative of the PSD transfer functions,
\begin{equation}
	\alpha(W)=-\dd (\log H)/\dd(\log W)~.
\label{eq:alphaW}
\end{equation}
In Figs.~\ref{fig:fig4X}, \ref{fig:fig5X}, and \ref{fig:fig6X} we show color plots of $\alpha(W)$ for the soma current ($\alpha^I(W)$),
current-dipole moment ($\alpha^p(W)$), and soma potential ($\alpha^V(W)$), respectively, both for cases with uncorrelated and correlated inputs. The depicted results are found by numerically evaluating
Eq.~\ref{eq:alphaW} based on the expressions for $H$ listed in Eqs.~\ref{eq:PSDIc}-\ref{eq:PSDVucs}.
Note that since our model is linear, the log-log derivative is independent of the amplitude $A$. Thus, with either completely correlated or completely uncorrelated input, the dimensionless parameters $B$, $L$, $\rho$ and $W$ span the whole parameter space of the model.
The 2D color plots in Figs.~\ref{fig:fig4X}-\ref{fig:fig6X} depict $\alpha$ as function of $W$ and $B$ for three different values of the
electronic length $L=l/\lambda$ ($L$=0.25, 1, and 4), i.e., spanning the situations from a very short dendritic stick ($L=0.25$) to a very long stick ($L=4$). Electrotonic lengths greater than $L=4$ produced plots that were indistinguishable by eye from the plots for $L=4$.
The thin black contour line denotes the transition between the low- and intermediate-frequency regimes ($\alpha=0.5 \alpha_\infty$), whereas the thick black contour line denotes the transition between the intermediate- and high-frequency regimes ($\alpha=0.9 \alpha_\infty$).

\

\centerline{FIG. \ref{fig:fig4X} AROUND HERE}

\

\subsubsection{Soma current}

Fig.~\ref{fig:fig4X} shows the slopes $\alpha$ of the PSD transfer functions for the soma current, $H^I$. The first row applies to correlated inputs ($H^I_\cc$) for \emph{all} values of $\rho_\s$ and $\rho_\dd$ as long as $\rho_\s \neq \rho_\dd$. This independence of $\rho=\rho_\s/(\rho_\s+\rho_\dd)$ is seen directly in the transfer functions in Eqs.~\ref{eq:PSDIc} and ~\ref{eq:PSDpc}.
(For the special case $\rho_\s=\rho_\dd$ there will be no net somatic current).
The plot in row 1 also applies to the case of uncorrelated current inputs onto the soma only ($H^I_{\uc,\s}$). That these particular PSD transfer functions have identical slopes are to be expected: correlated result pertains also to the special case $\rho_\dd=0$ for which all input is onto the soma, and changing from correlated to uncorrelated  current inputs onto the soma
will only change the overall amplitude of the resulting soma current, not the PSD slope.

The first row of Fig.~\ref{fig:fig4X} illustrates how the slope $\alpha$ approaches the asymptotic value $\alpha^I_\infty=1$ for correlated input ($\rho_\s \neq \rho_\dd$) (and uncorrelated input onto the soma) for high frequencies, see Table~\ref{tab:series_parameters}. It also shows that this asymptotic value is reached for lower frequencies when $B=d_\s^2/(d\lambda)$ is large, i.e., when the soma area is large compared to the effective area $\lambda d$ of the dendrite.  Row 2 correspondingly shows how $\alpha$ for large frequencies approaches the asymptotic value of $\alpha^I_\infty=1/2$ (row~2) for uncorrelated input uniformly spread over the dendrite. For the case depicted in row 3, i.e., uncorrelated input onto \emph{both} the soma and dendrite with $\rho_\s=\rho_\dd$, the asymptotic high-frequency expression is seen to eventually be dominated by the lowest power, i.e., $\alpha \approx \alpha^{\mathrm{all},I}_\infty=1/2$.

The \emph{lf} regime, that is, the area to the left of the thin contour line, is seen to be quite substantial in Fig.~\ref{fig:fig4X}, and is also highly dependent on $B$. For the default parameters, depicted by the white horizontal line, the left column in Fig.~\ref{fig:fig4X} shows that the \emph{lf} regime extends up to much more than 100~Hz for compact neurons ($L$=0.25), and even for $L=1$ and $L=4$ (two rightmost columns) the \emph{lf} regimes are substantial. (For our default membrane time constant of 30~ms, 100~Hz corresponds to the middle vertical white line in the panels.) Such a prominent \emph{lf} regime was also seen for the pyramidal neuron in Fig.~\ref{fig:fig2X} where the normalized PSD for the somatic membrane current with uncorrelated input was almost constant up to 1000~Hz.

It is also interesting that in some situations the soma current is band-pass filtered with respect to the input currents. This is especially seen in Fig.~\ref{fig:fig4X} for intermediate ($L=1$) and long ($L=4$) sticks with uncorrelated dendritic input currents (row~2), where the substantial dark blue area  represents a band of negative $\alpha$-values which is turning positive for higher frequencies, and the PSD thus is band-pass filtered around the frequencies corresponding to $\alpha=0$. For the higher frequencies within the frequency interval typically recorded in experiments (up to a few hundred hertz), Fig.~\ref{fig:fig4X} shows that one could expect some low-pass filtering for the intermediate and long sticks ($l > \lambda$), in particular if the current input is (i) predominantly onto the soma or (ii) correlated, and the neuron has a large value of $B$.  However, as indicated by Fig.~\ref{fig:fig2X}, this effect may be very small for pyramidal neurons.

\

\centerline{FIG. \ref{fig:fig5X} AROUND HERE}

\

\subsubsection{Current-dipole moment / EEG contribution}

Fig.~\ref{fig:fig5X} shows corresponding slope plots of the PSD for the current-dipole moment, $H^p$, i.e., the single-neuron contribution to the EEG.
The panels are organized as for the soma current in Fig.~\ref{fig:fig4X}, and as for the soma current we observe that
for high frequencies $\alpha$ approaches the asymptotic value $\alpha^p_\infty$=2 for the cases with either correlated input ($\rho_\s \neq \rho_\dd$) or uncorrelated input onto the soma only (row 1), see Table~\ref{tab:series_parameters}. Further, for the case with uncorrelated input on the dendrites, $\alpha$ is seen to approach the predicted $\alpha^{\mathrm{all},p}_\infty=1.5$ (rows 2 and 3).

Moreover, as for the soma current the \emph{lf} regime is seen to be large for compact neurons ($L$=0.25). For such neurons one would thus expect very little filtering within the frequency interval typically recorded for the EEG, typically up to 100 or 200~Hz (middle vertical white line in panels). For less compact neurons ($L$=1 and 4), the filtering is, however, seen to be substantial also within the frequency interval from 10 to 100~Hz, even for low values of $B$.
This filtering is seen to be even more prominent for the pyramidal neuron in Fig.~\ref{fig:fig2X}, suggesting that the filtering could be of considerable importance for the large pyramidal neurons in human cortex thought to dominate human EEG.

The \emph{if} regime is seen to be quite narrow in all panels in Fig.~\ref{fig:fig5X}, implying that the PSD has a quite abrupt transition to the \emph{hf} regime where the slope is quite constant and close to its asymptotic values $\alpha_\infty^p$. The pyramidal neuron receiving uncorrelated input in Fig.~\ref{fig:fig2X}, however, is seen to obey an approximate power-law with $\alpha^p$ of only about 1.25 at 1000~Hz. This is not within the range defined here as the \emph{hf} regime, i.e., $\alpha \geq 0.9 \alpha_\infty^p=1.35$, but rather within the upper range of the \emph{if} regime.

\

\centerline{FIG. \ref{fig:fig6X} AROUND HERE}

\

\subsubsection{Soma potential}

In Fig.~\ref{fig:fig6X} the slopes $\alpha$ of the PSD of the soma potential are shown. Unlike $H^I_\cc$ and $H^p_\cc$, the PSD transfer function $H^V_\cc$ for the soma potential with correlated input currents varies with $\rho=\rho_\s/(\rho_\s+\rho_\dd)$,
and is also non-zero for $\rho_\s=\rho_\dd$, cf. Eq.~\ref{eq:PSDVc}. More panels are thus needed to describe the model predictions properly:
Row~1 corresponds to correlated input onto the dendrite only ($H^V_\cc (\rho_\s=0)$), row~2 corresponds to somatic input only, either correlated ($H^V_\cc(\rho_\dd=0)$) or uncorrelated ($H^V_{\uc,\s}$), while row~3 corresponds to uncorrelated dendritic input ($H^V_{\uc,\dd}$). The two bottom rows correspond to homogeneous input onto the whole neuron, i.e., $\rho_\dd=\rho_\s$, with uncorrelated input in row~4 and correlated input in row~5.

The different panels of Fig.~\ref{fig:fig6X} display quite varied PSD slopes for the various scenarios of input current.
Row~1 shows that for correlated input solely onto the dendrite, $\alpha$ is quite close to the asymptotic value $\alpha_\infty^V$=3
(cf. Table~\ref{tab:series_parameters}) for modest frequencies, even for the compact neuron with $L=0.25$. The narrow \emph{if} region and large power-law exponent $\alpha$ in row~1 makes this case quite different from the results depicted in the other panels. With input instead onto the soma only (row~2), for example, a completely different slope pattern is observed: for compact neurons ($L=0.25$) the log-log slope of the PSD is seen to have regions with a positive double derivative (concave slope), with the consequence that the \emph{if} regime is divided into two distinct frequency regions with an intermediate \emph{hf} interval.

Row~3 depicts the case with uncorrelated input onto the dendrites. Qualitatively the results resemble the case with correlated dendritic inputs in row 1, except that here $\alpha$ approaches the asymptotic values $\alpha^V_\infty=2.5$ (cf.  Table~\ref{tab:series_parameters}), rather than 3. For the non-compact neurons ($L=1$ and $L=4$) the default parameters give an \emph{if} region for uncorrelated dendritic input which goes up to almost 100~Hz. However, the thick contour line illustrates that the transition to the \emph{hf} regime is highly dependent on the values $B$, and a slightly larger $B$ is seen to substantially lower the transition frequency to the \emph{hf} regime.

With uncorrelated input homogeneously distributed over the whole neuron, i.e., $\rho_\s=\rho_\dd$ (row~4), we observe a similar pattern of power-law exponents as for somatic input only (row~2). Thus the contribution from the soma for which $\alpha^V_\infty=2$, dominates the contribution from the dendritic inputs where $\alpha^V_\infty$=2.5. Another observation is that for the non-compact neurons ($L$=1 and 4)
the \emph{if} regime is wide for a large range of $B$ values. For the default parameters corresponding to $B$=0.2 we observe that the \emph{if} interval stretches from less than 10~Hz to almost 1000~Hz.

For the last example case in row~5 with correlated input spread homogeneously onto the whole neuron ($\rho_\s=\rho_\dd$) we observe that $\alpha$ is independent of the parameter $B$. For homogenous correlated input the whole neuron is iso-potential and corresponds to a single-compartment neuron with zero dipole moment and zero net membrane current,  as reflected in the vanishing amplitudes of $A^I_\cc$ and $A^p_\cc$ in Table~\ref{tab:series_parameters}. In this special case the spatial extension of the dendritic stick will not affect the filtering properties of the neuron, and the PSD transfer function can be expressed as a simple Lorentzian, i.e., $\left. H^V_\cc \right|_{\rho=0.5} \propto 1/(1+W^2)$. The slope $\alpha$ is thus solely determined by the membrane time constant $\tau_\m$ hidden within the dimensionless frequency $W=2 \pi f \tau_\m$.

\subsection{D. PSDs for varying biophysical parameters for ball and stick neuron}

The 2D color plots in Figs.~\ref{fig:fig4X}--\ref{fig:fig6X} depicting the slopes $\alpha$ of the PSDs of the transfer functions $H(f)$, give a comprehensive overview of the power-law properties of the ball and stick model as they are given in terms of the three key dimensionless parameters $W= \omega \tau_m=2 \pi f \tau_\m$, $B=d_s^2/d\lambda$, and $L=l/\lambda$. To get an additional view of how the
model predictions depend on biophysical model parameters, we plot in Figs.~\ref{fig:fig7X} and \ref{fig:fig8X} PSDs, denoted $S(f)$, for a range of model parameters for the soma current, current-dipole moment and soma potential when the neuron receives homogeneous white-noise current input across the dendrite and/or the soma. We focus on biophysical parameters that may vary significantly from neuron to neuron: the dendritic stick length $l$, the specific membrane resistance $R_\m$, the dendritic stick diameter $d$, and the soma diameter $d_\s$. The specific membrane resistance may not only vary between neurons, but also between different network states for the same neuron~\cite{Waters2006,Destexhe2012}.

To predict PSDs $S(f)$ of the various measurements, and not just PSDs of the transfer functions $H(f)$, we also need to specify numerical values for the current-input densities $\rho_\dd$ and $\rho_\s$ (and not only the ratio $\rho=\rho_\s/(\rho_\s+\rho_\dd)$), as well as the magnitude of the PSDs of the current inputs. These choices will only affect the magnitudes of the predicted PSDs, not the power-law slopes. As the numerical values of the slopes predicted by the present work suggest that channel noise from intrinsic membrane conductances rather than synaptic noise dominates the observed noise in experiments (see Discussion), we gear our choice of parameters towards intrinsic channel noise. We first assume the input densities $\rho_\dd$ and $\rho_\s$ (when they are non-zero)
to be 2 $\mu$m$^{-2}$, in agreement with measurements of the density of the large conductance calcium-dependent potassium (BK) channel \cite{Benhassine2005}. Next we assume the magnitude of PSD of the white-noise current input to be $s(f)$=const.=1~fA$^2$/Hz. This choice for $s$ gives magnitudes of predicted PDSs of the soma potential, assuming uncorrelated current inputs, in rough agreement with what was observed in \cite{Diba2004}, i.e., about 10$^{-3}$--10$^{-2}$~mV$^2$/Hz for low frequencies.

Figs.~\ref{fig:fig7X} and \ref{fig:fig8X} show PSDs for uncorrelated and correlated input currents, respectively. A first observation is that the predicted PSD magnitudes are typically orders of magnitude larger for correlated inputs, than for uncorrelated inputs. With the present choice of parameters, the cases with correlated inputs predict PSDs for the soma potential and soma current much larger than what is seen in experiments~\cite{Diba2004,Jacobson2005,Yaron-Jakoubovitch2008}. A second observation is that variations in the dendritic stick length (first column in Figs.~\ref{fig:fig7X}-\ref{fig:fig8X}) and membrane resistance (second column) typically have little effect on the PSDs at high frequencies, but may significantly affect the cut-off frequencies, i.e., the frequency where the PSD kinks downwards. This may be somewhat counterintuitive, especially that the PSDs for the current-dipole moment are independent of stick length $l$ as one could think that a longer stick gives a larger dipole moment. For the ball and stick neuron, however, this is not so: input currents injected far away from both boundaries (ends) of a long stick will not contribute to any net dipole moment, as the input current will return symmetrically on both sides of the injection point and thus form a quadrupole moment.
This symmetry is broken near the ends of the stick: for uncorrelated input a local dipole is created at each endpoint;
for correlated input the dendrite will be iso-potential near the distal end of the stick, while a local dipole will arise at the somatic end if $\rho_\dd \ne \rho_\s$. Note though that this is expected to be different for neurons with realistic dendritic morphology, since the dendritic cables typically are quite asymmetric due to branching and tapering.

The effects of varying the dendritic stick diameter and soma diameter are quite different (cf., two rightmost columns in
Figs.~\ref{fig:fig7X}--\ref{fig:fig8X}). Here both the magnitudes and the slopes of the high-frequency parts are seen to be significantly
affected. On the other hand, the cut-off frequency is seen to be little affected  when varying the soma diameter $d_\s$, in particular for the current-dipole ($S^p$) and soma potential ($S^V$) PSDs. (Note that for the case with homogeneous correlated input, $\rho_\s=\rho_\dd$  (row 4 in Fig.~\ref{fig:fig8X}), the ball and stick model is effectively reduced to a single-compartment neuron for which the PSD is independent of $d$ and $d_\s$.)


In Figs.~\ref{fig:fig4X}--\ref{fig:fig6X} regions in the log-log slope plots were observed to have positive double derivatives, i.e., concave curvature. The effect was particularly prevalent for the soma potential transfer function $H^V$ in the case of short dendritic sticks ($L=0.25$) with dominant current input to the soma. This feature is also seen in the corresponding 'soma-input' curves (bottom rows of Figs.~\ref{fig:fig7X}--\ref{fig:fig8X}), also for non-compact sticks, i.e., for the default value $l$=1~mm ($L$=1).

%

\section{IV. Discussion}

\subsection{A. Summary of main findings}
In the present work we have taken advantage of the analytical tractability of the ball and stick neuron model to obtain general expressions for the power spectral density (PSD) transfer functions for a set of measures of neural activity: the somatic membrane current, the current-dipole moment (corresponding to the single-neuron EEG contribution), and the soma potential. With homogeneously distributed input currents both onto the dendritic stick and with the same, or another current density, onto the soma we find that all three PSD transfer functions, relating the PSDs of the measurements to the PSDs the noisy inputs currents, express asymptotic high-frequency $1/f^\alpha$ power laws. The corresponding power-law exponents are analytically identified as $\alpha^{I}_\infty=1/2$ for the somatic membrane current, $\alpha^{p}_\infty=3/2$ for the current-dipole moment, and $\alpha^{V}_\infty=2$ for the soma potential. These power-law exponents are found for arbitrary combinations of uncorrelated and correlated noisy input current (as long as both the dendrites and the soma receive some uncorrelated input currents).

The significance of this finding goes beyond neuroscience as it demonstrates how $1/f^\alpha$ power laws with a wide range of values for the power-law exponent $\alpha$ may arise from a simple, linear physics equation. We find here that the cable equation describing the electrical properties of membranes, transfers white-noise current input into 'colored' $1/f^\alpha$-noise where $\alpha$ may have any half-numbered value within the interval from $1/2$ to 3 for the different measurement modalities. Intuitively, the physical underpinning of these novel power laws is the superposition of numerous low-pass filtered contributions with different cut-off frequencies (i.e., different time constants) due to the different spatial positions of the various current inputs along the neuron. As our model system is linear, the results directly generalize to any colored input noise, i.e., transferring  $1/f^\beta$ spectra of input currents to $1/f^{\beta+\alpha}$ output spectra.

\subsection{B. Comparison with power laws observed in neural recordings}

Our ball and stick model expressions for the PSDs cover all frequencies, not just the high frequencies where the power-law behavior is seen. When comparing with results from neural recordings, one could thus envision to compare model results with experimental results across the entire frequency spectrum. However, the experimental spectra will generally be superpositions of contributions from numerous sources, both from synapses \cite{Destexhe2012} and from ion channels~\cite{Diba2004}. These various types of input currents will in general have different PSDs, i.e., different $s(f)$.
A full-spectra comparison with our theory is thus not possible without specific assumptions about the types and weights of the various noise contributions, information which is presently not available from experiments. However, the presence of power-law behavior at high frequencies implies that a single noise process (or several noise processes with identical power-law exponent) dominates the others in this frequency range.

In in vivo experiments from neocortical neurons where PSDs for frequencies up to 1000~Hz or more have been used to estimate power-law exponents, the soma potential has typically been seen to express power laws with $\alpha_{\mathrm{exp}}^V$ close to 2.6~\cite{Destexhe2003,Rudolph2005,Bedard2008}.
In an analogous experiment in hippocampal cell culture where the PSD for frequencies up to 500~Hz was measured, a value of $\alpha_{\mathrm{exp}}^V$ of about 2.4 was estimated~\cite{Diba2004}.
For the soma current the results are fewer, but for voltage-clamped neurons in hippocampal cell cultures a power-law with $\alpha_{\mathrm{exp}}^I=1.1$ was seen in the high-frequency end of the PSD recorded up to 500~Hz~\cite{Diba2004}. For the pyramidal neuron depicted in Fig.~\ref{fig:fig2X} we correspondingly found $\alpha^V$=1.61 and $\alpha^I$=0.15 for the PSD of the transfer functions for \emph{uncorrelated} current inputs. Thus if these uncorrelated input current sources themselves have a pink ($1/f$, i.e., $\beta$=1) power-law dependence of the PSD in the relevant frequency range, the power-law exponents of the model PSDs become $\alpha^V$+$\beta$=1.61+1=2.61 and $\alpha^I+\beta$=1.15, intriguingly close to the experimental observations. Note that while these model results pertain for a particular choice of model parameters for the pyramidal neuron, the results shown in Fig.~\ref{fig:fig7X} for the ball and stick neuron implies that moderate changes in the model parameters will yield modest changes in predicted power-law exponents.

For the EEG, only experimental findings up to frequencies of 100~Hz are available, and here estimated power laws have exhibited a large variation in power-law exponents with $\alpha_{\mathrm{exp}}$'s varying between 1 and 2~\cite{Freeman2003}. If uncorrelated pink-noise ($1/f$) input currents are assumed
also here, the pyramidal neuron results in Fig.~\ref{fig:fig2X} imply $\alpha^p+\beta$=2.25, i.e., the single-neuron contribution to the EEG exhibits a power law with an exponent somewhat above the typical value for macroscopic recordings. Note, however, that even for pink-noise input, shorter dendritic sticks may imply power-law exponents as small as $\alpha^p+\beta$=1 for the single-neuron EEG contribution within the lower frequency range typically probed in EEG recordings (Fig.~\ref{fig:fig5X}, \cite{Freeman2003}). Note also that the link between our model predictions and putative EEG power laws is more tenuous as it involves the additional assumption that network dynamics do not affect the observed high-frequency power laws, i.e., the high-frequency tail of the input current PSD $s(f)$ of the neurons giving the dominant contributions to the EEG signal.

On balance we think the comparison with presently available experiments supports an hypothesis that the observed microscopic (soma potential, soma current) power laws, or more precisely 'apparent' power laws, stem from (i) uncorrelated membrane current sources which are (ii) homogeneously distributed across the neural membranes and (iii) each have a pink ($1/f$) noise distribution. Further, while experimental data presently are scarce, this hypothesis may also explain the presence of power laws in EEG recordings.

Note, however, that power-law exponents alone are not sufficient to uniquely determine whether the dominant inputs are correlated or uncorrelated. As seen for the 'infinite-frequency' power-law
exponents $\alpha_\infty$ in Table~\ref{tab:series_parameters} and Figs.~\ref{fig:fig4X}-\ref{fig:fig8X},
$\alpha$'s are equal to or larger for correlated inputs than for uncorrelated inputs for our ball and stick neuron; the typical difference for $\alpha_\infty$ being 1/2. Thus correlated current inputs with power-law PSDs with an exponent $\beta$ of about 1/2 (rather than the pink-noise value of $\beta$=1) would give about the same power-law exponent ($\alpha$+$\beta$) in the various measurements. Note also that since the power-law exponents $\alpha$ with uncorrelated inputs are generally smaller than for correlated inputs, the uncorrelated contributions will in principle always dominate for sufficiently high frequencies. However, the contribution from correlated current inputs scales differently with the number density of input currents than for uncorrelated inputs: the PSD grows as the \emph{square} of the input densities ($\rho_s$, $\rho_d$) for correlated inputs, while it grows only \emph{linearly} with these input densities for uncorrelated inputs. Thus in experimental settings the relative contributions from correlated and uncorrelated current inputs will depend on the size of these densities as well as the value of the coherence $c$, parameters which cannot be expected to be universal, but rather depend on the biophysical nature of the underlying current noise source. It is thus difficult to a priori assess whether the noise spectra are dominated by correlated or uncorrelated input. As argued on biophysical grounds in the next subsection, however, we think that the explanation assuming uncorrelated inputs with pink noise is more likely.

\subsection{C. Origin of noise}

Our supposition that homogeneously distributed, uncorrelated pink-noise current input, i.e. $s=s(f) \sim 1/f$ input, may underlie the observed power-law behavior in the soma current, soma potential, and possibly also EEG spectra, applies regardless of the biophysical nature of the underlying noisy currents. Nevertheless, we will in the following discuss the origin of the noise which in previous modeling studies have been assumed to stem from synapses~\cite{Bedard2008,Yaron-Jakoubovitch2008}, intrinsic ion channels \cite{Diba2004}, or a combination of the two \cite{Jacobson2005}. Our argument presented below that the high-frequency power-law behavior predominantly stems from channel noise follows from (i) the observed lack of change in power-law exponents when synaptic inputs are blocked \cite{Jacobson2005}, (ii) the previous experimental observation of pink noise in certain ion channels \cite{Derksen1966,Poussart1971,Fishman1973,Siwy2002}, and (iii) the difficulty of reconciling pink noise from synapses with prevailing synapse models \cite{Yaron-Jakoubovitch2008,Bedard2008}. (Note, however, that this does not imply that channel noise rather than synaptic noise dominates the PSD at lower frequencies.)

Most detailed studies of neuronal noise spectra have focused on the soma potential \cite{Destexhe2003,Diba2004,Rudolph2005,Jacobson2005,Yaron-Jakoubovitch2008}, where interestingly the same power-law exponent of about $\alpha_V=2.6$
have been observed both in \emph{in vivo}~\cite{Destexhe2003,Rudolph2005,Bedard2008} and
\emph{in vitro} conditions~\cite{Diba2004,Jacobson2005,Yaron-Jakoubovitch2008}. The most parsimonious explanation is that the same noise process dominates under both conditions, i.e., that the higher spiking activity in \emph{in vivo} conditions than in \emph{in vitro} conditions is not the key process underlying the observed power law. This suggests that the dominant mechanism rather is noise stemming from intrinsic ion channels. This conjecture is supported by the observation that the slope of the soma potential power law in rat neocortical slices was not affected by application of synaptic blockers (DNQX, gabazine)~\cite{Jacobson2005}. (However, the synaptic blockers reduced the overall amplitude of the PSD, which would imply that a secondary effect is involved, e.g., that blocking of synaptic inputs indirectly affects the amplitude of the ion-channels noise by, for example, changing the intracellular calcium concentration.)

Further, it has been difficult to account for $1/f$ input spectra in model studies based on assuming a synaptic origin. In \cite{Yaron-Jakoubovitch2008} and \cite{Bedard2008} synapses were spread evenly across dendrites of morphologically reconstructed neurons and were activated by presynaptic spike trains assumed to have Poissonian distributions (cf. Fig.~\ref{fig:fig1X}). With current-based exponential synapses, the PSD of the current noise source will then
have the form of a Lorentzian, i.e., $s(f)\propto1/(1+(2 \pi f \tau_\s)^2)$, where $\tau_\s$ is the synaptic time constant. For high frequencies this implies $s \sim 1/f^2$
($\beta$=2),
cf. results for Brownian ($1/f^2$) input in right column of Fig.~\ref{fig:fig2X}. As previous studies also found that this implies a too large value for the soma potential power-law exponent several approaches has been suggested to compensate for this: \cite{Boustani2009} showed that network correlations due to delay distributions can give non-Poissonian pre-synaptic spike-train statistics and thus change the power-law exponent. Alternatively, small synaptic time constants ($\tau_\s=2$~ms in \cite{Yaron-Jakoubovitch2008}) will give a higher cut-off frequency ($f=1/2 \pi \tau_\s$) for the transition to the high-frequency power-law regime. If this cut-off frequency is in the upper range of the recorded frequency interval, $s(f)$ will essentially be independent of frequency (i.e., white) and apparent soma-potentials power laws with smaller exponents can be obtained. In \cite{Bedard2008} it was instead suggested that a non-ideal membrane capacitance could have a compensatory effect.

In contrast, several recordings of PSDs of the intrinsic channel noise in potassium channels have shown $1/f$ scaling \cite{Derksen1966,Poussart1971,Fishman1973,Siwy2002}, i.e., exactly the type of pink input-current noise spectrum required for our model prediction to be in accordance with the experimentally observed power-law exponents in the PSDs of the soma current, soma potential, and EEG. If the power law indeed stems from intrinsic channel noise, uncorrelated rather than correlated current sources are expected. Again this agrees with predictions from comparing our model with experiments. Further, it is tempting to speculate on what particular type of ion channel could give rise to the observed power-law spectra. Several experiments have hinted that potassium channels may be important sources of membrane noise \cite{Derksen1966,Poussart1971,Fishman1973,Diba2004}, and of those a natural candidate is the BK ('big') potassium channel which has a large single-channel conductivity and thus the potential for large current fluctuations.

\subsection{D. Power laws for local field potentials (LFP) and ECoG signals}

Power laws have also been reported in recordings of extracellular potentials inside (local field potential; LFP) and at the surface of cortex (electrocorticography; ECoG).
However, the reported power-law exponents vary a lot, with $\alpha_\mathrm{exp}$'s between 1 and 3 for
LFPs~\cite{Bedard2006,Milstein2009,Martinez2009,Baranauskas2012} and between 2 and 4 for ECoG
~\cite{Freeman2000,Freeman2009,Miller2009,He2010,Zempel2012}. From a modeling perspective the single-neuron contribution to putative power-law exponents for these signals is more difficult as, unlike the EEG signal, the single-neuron contributions are not determined only by the current-dipole moment: dominant contributions to these signals will come from neurons close to the electrode (typically on the order of hundred or a few hundred micrometers~\cite{Linden2011}), so close that the far-field dipole approximation relating the current-dipole moment directly to the contributed extracellular potential~\cite{Hamalainen1993}
is not applicable~\cite{Linden2011}.

A point to note, however, is that it may very well be that power laws observed in the LFP or ECoG are dominated by other current sources than the power laws observed in the EEG spectra: As observed in
\cite{Linden2011,Leski2013} (see also \cite{Einevoll2013}) the LFP recorded in a cortical column receiving correlated synaptic inputs can be very strong, and it is thus at least in principle conceivable that power laws in the LFP may stem from synaptic inputs from neurons surrounding the electrode, whereas the EEG signal, which  picks up contributions from a much larger cortical area, may be dominated by uncorrelated noise from ion channels. Further, the soma potential and soma current of each single neuron may also still be dominated by uncorrelated channel noise, even if the the LFP is dominated by correlated synaptic activity. This is because correlated synaptic inputs onto a population of neurons add up constructively in the LFP, whereas the uncorrelated inputs do not~\cite{Linden2011,Leski2013}. For single-neuron measures such as the soma potential and soma current there will be no such population effects, and the uncorrelated inputs will more easily dominate the power spectra.

As a final comment it is interesting to note that in the only reported study we are aware of for the frequency range 300-3000~Hz, the PSD of the LFP exhibited a power law with a fitted exponent of $\alpha$=1.1~\cite{Martinez2009}. This is very close to what would be predicted if the LFP was dominated by the soma current from uncorrelated (pink) noise sources: In Table~\ref{tab:series_parameters} we see that the 'infinite-frequency' power-law exponent for the transfer function from dendritic current inputs to soma current is $\alpha_\infty^{I}=0.5$. With a pink ($1/f$) PSD of the input noise current, the 'infinite-frequency' prediction for the soma current exponent will thus be 1.5. This is already fairly close to the experimental observation of 1.1. Further, from Fig.~\ref{fig:fig4X} it follows that the apparent power-law coefficient for the transfer-function power law may be somewhat smaller than 0.5 in the frequency range of interest, suggesting that the agreement between experiments and model predictions assuming uncorrelated noise may be even better. If so, it may be that the LFP power spectra are dominated by synaptic inputs for frequencies below a few hundred hertz (with rapidly decaying LFP contributions with increasing frequency, i.e., higher power-law exponents in accordance with \cite{Bedard2006,Milstein2009,Baranauskas2012}), while uncorrelated inputs, and thus power laws with  smaller exponents, dominate at higher frequencies.

\subsection{E. Passive approximation }

In the present analysis we have modeled the membranes of somas and dendrites as simple passive linear (RC) circuit elements. This implies a strictly linear response to the current inputs, allowing for the present frequency-resolved (Fourier) analysis. The present results also serve as a starting point for the exploration of non-linear effects, for example due to active membrane conductances. Close to the resting potential of the neuron, the active conductances can be linearized, and the neuron dynamics can be described by linear theory with quasi-active membrane modeled by a combination of resistors,
capacitors and inductors \cite{Chandler1962,Koch1984}. These extra circuit elements will change the PSD. For example, the inductor typically introduces a resonance in the system. In Koch~\cite{Koch1984} the impedance for this 'quasi-active' membrane was however found to coincide with the impedance for a purely passive membrane for frequencies above 200~Hz,
implying that the predicted high frequency power laws will be about the same.
This is in accordance with experimental results from neocortical slices, where blocking of sodium channels were shown mainly to affect the soma potential PSD for frequencies below 2~Hz~\cite{Jacobson2005}. Nevertheless, the investigation of the role of active conductance on PSDs is a topic deserving further investigations.

\subsection{F. Concluding remarks}

A key conclusion from the present work is that the power-law predictions from our models are in close agreement with experimental findings for the soma potential and the soma current provided the transmembrane current sources are assumed to be (i) homogeneously distributed throughout the whole neuron, (ii) uncorrelated, and (iii) have a pink ($1/f$) noise distribution. It should be stressed that we do not argue against synaptic noise being a major component underlying neural noise spectra; the importance of synaptic inputs in setting the noise level has been clearly demonstrated,  for example by the large difference in membrane potential fluctuation between in vivo and in vitro preparations~\cite{Destexhe2003,Destexhe2012}. We rather suggest that the power-law behavior seen at the high-frequency end of these noise spectra are dominated by intrinsic channel noise, not synaptic noise.

We also speculate that potassium channels with inherent noisy current with PSDs following a $1/f$ distribution in the relevant frequency range, underlie the observed power laws, and the BK channel is suggested as a main contributor.  If future experiments indeed confirm that the BK channel is a dominant source of membrane noise, this may have direct implication of the understanding several pathologies. Not only has the BK channel been implicated as a source of increased neural excitability~\cite{Gerkin2010} and epilepsy~\cite{Du2005}, but also disorders such as schizophrenia \cite{Zhang2006}, autism and mental retardation \cite{Laumonnier2006} have been linked to the BK channel through a decrease in its expression \cite{Lee2010}.

\section{Acknowledgments}
We thank Eivind Norheim for useful feedback on the manuscript and for testing the reproducibility of Figs.~\ref{fig:fig4X}-\ref{fig:fig6X}. This work was partially funded by the Research Council of Norway (eVita [eNEURO], NOTUR,ISP) and EU Grant 269921 (BrainScaleS).


\bibliography{Pettersen2013_bib_v2}

\clearpage
\section{Figure Legends}

\

\begin{figure}[!ht]
\begin{center}
\includegraphics[width=6in]{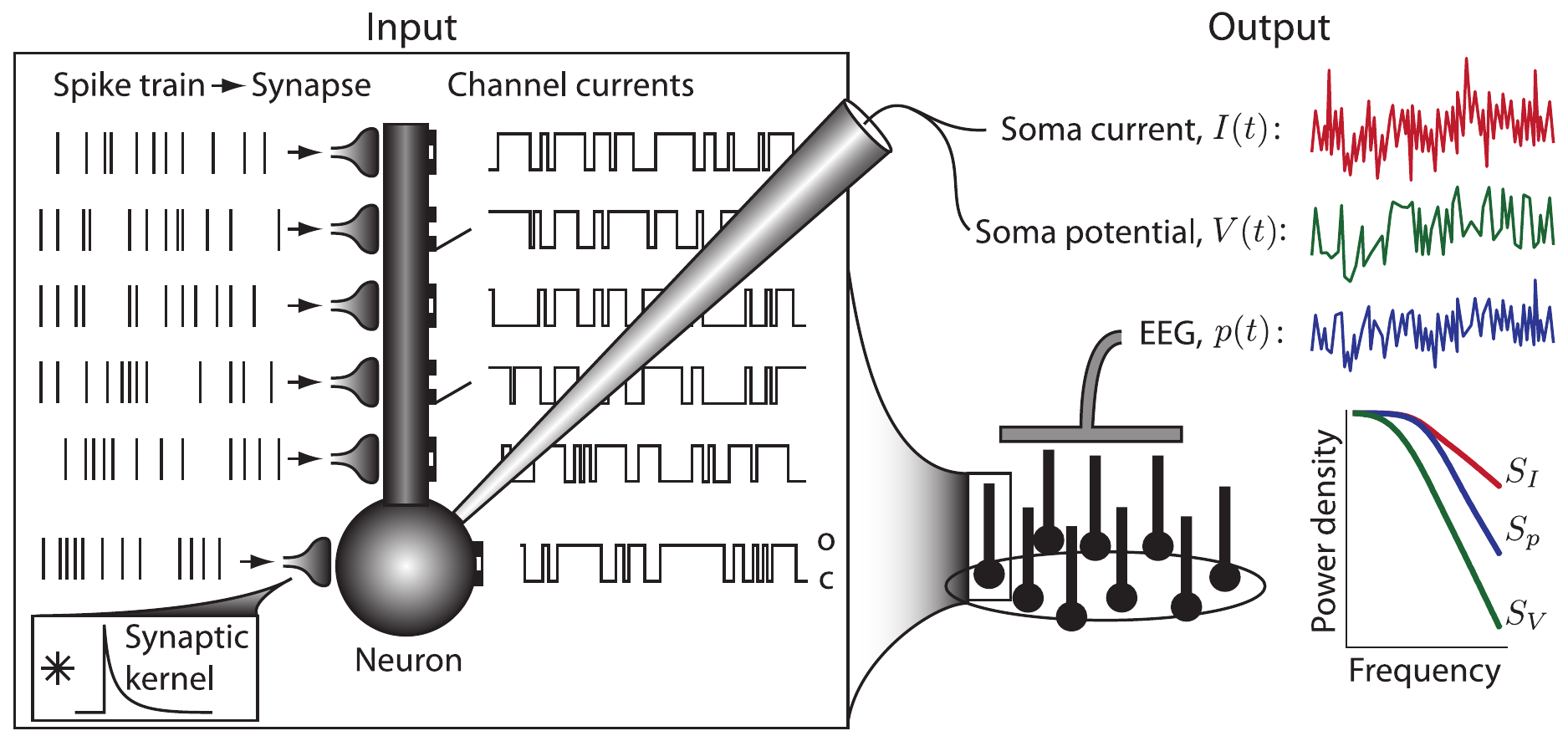}
\end{center}
\caption{
}
\label{fig:fig1X}
\end{figure}

\noindent
{\bf Figure~\ref{fig:fig1X}}: Schematic illustration of the input-output relationship between transmembrane currents (input) and the different measurement modalities (output). The transmembrane currents are illustrated by synaptic currents and channel currents. 
A synaptic current is commonly modeled by means of exponentially decaying functions
(synaptic kernel) triggered by incoming spike trains, whereas a channel current typically is modeled by a channel switching between an open state (o), letting a current with constant amplitude through the channel, or a closed state (c). The input currents are filtered by the neuronal cable, resulting in a low-pass filtered output current in the soma with a power spectral density (PSD) designated $S_I$. The PSDs of the other measurement modalities studied here, i.e., the soma potential ($S_V$) and the current-dipole moment giving the single-neuron contribution to the EEG ($S_p$), are typically even more low-pass filtered, as illustrated by the PSDs plotted in the lower right panel.

\

\begin{figure}[!ht]
\begin{center}
\includegraphics[width=6in]{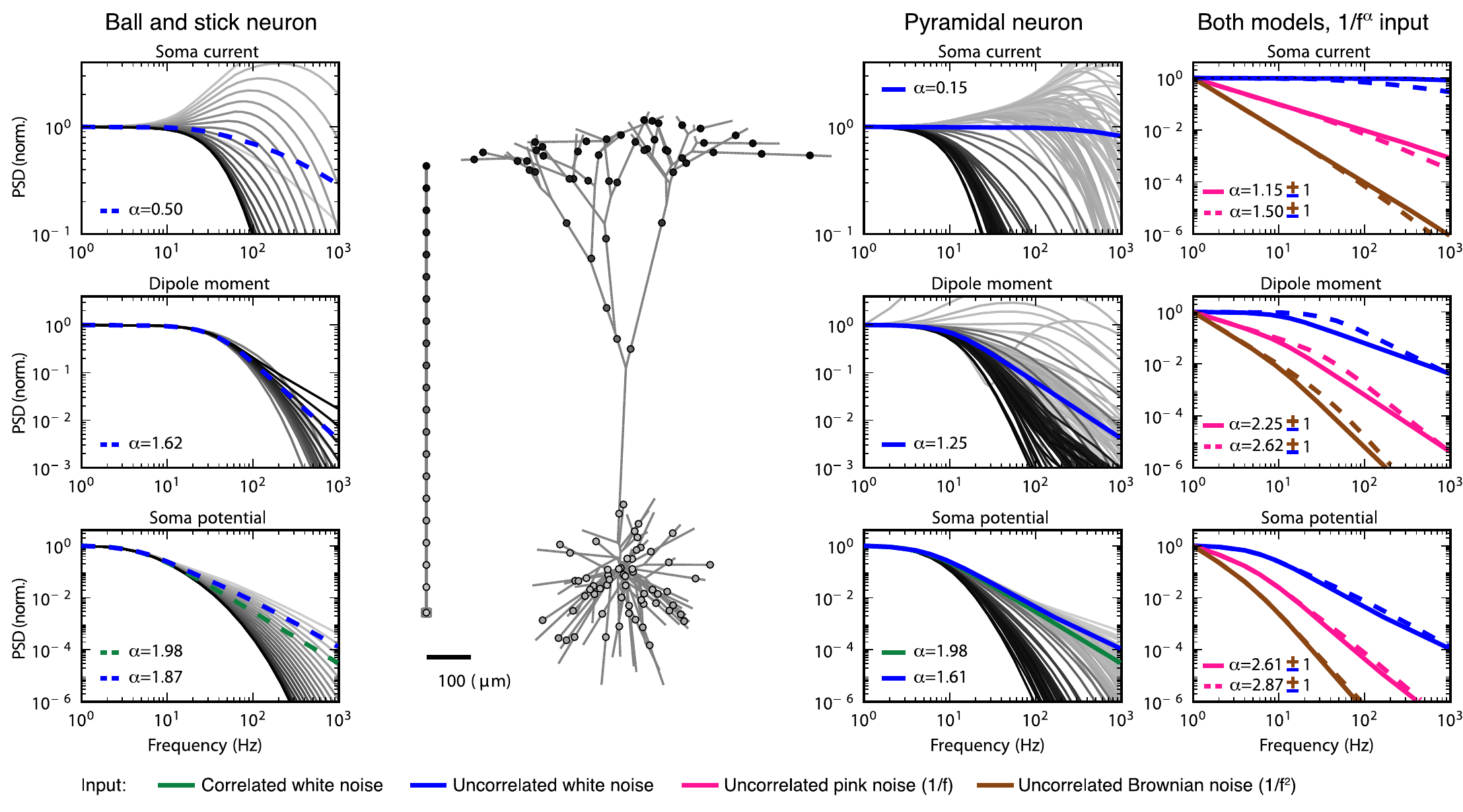}
\end{center}
\caption{
}
\label{fig:fig2X}
\end{figure}

\noindent
{\bf Figure~\ref{fig:fig2X}:} Normalized power spectral densities (PSDs) for the soma current, the current-dipole moment (i.e., EEG contribution) and the soma potential for a ball and stick neuron and a pyramidal neuron.  A homogeneous density of noisy input currents is applied throughout the neural membrane. Columns~1 (ball and stick neuron) and 2 (pyramidal neuron) show PSDs for white-noise input, the blue and green lines correspond to uncorrelated and correlated input currents, respectively.  Note that there is no green line in the two upper rows, since a homogeneous density of correlated inputs throughout the neuron gives no net soma current or dipole moment. An ensemble of PSDs from 20 single input currents for the ball and stick neuron and 107 single input currents for the pyramidal neuron is shown in grey. The results for the most distal synapses are shown in dark grey and the results for the proximal synapses in light grey, corresponding to the color shown in the filled circles at the respective neuron morphology (between columns 1 and 2). Column~3 illustrates how different 
power-law spectra of the input currents change the output PSDs: the blue, pink and brown lines express the PSD for uncorrelated white (constant), pink ($1/f$) and Brownian noise input ($1/f^2$), respectively. The values of $\alpha$ in legends denote estimated power-law exponents  at 1000~Hz, i.e., the negative discrete log-log derivative, $-\Delta(\log S)/\Delta(\log f)$. In the rightmost column the values of $\alpha$ correspond to pink noise input, for Brownian noise input and white-noise input the values are '+1' and '-1' with respect to the pink input, respectively, as indicated by the brown '+' and the blue '-'. The ball and stick neuron was simulated with 200 dendritic segments, while the pyramidal neuron was simulated with 3214 dendritic segments. Broken lines correspond to the ball and stick neuron, whole lines to the pyramidal neuron.

\

\begin{figure}[!ht]
\begin{center}
\includegraphics[width=6in]{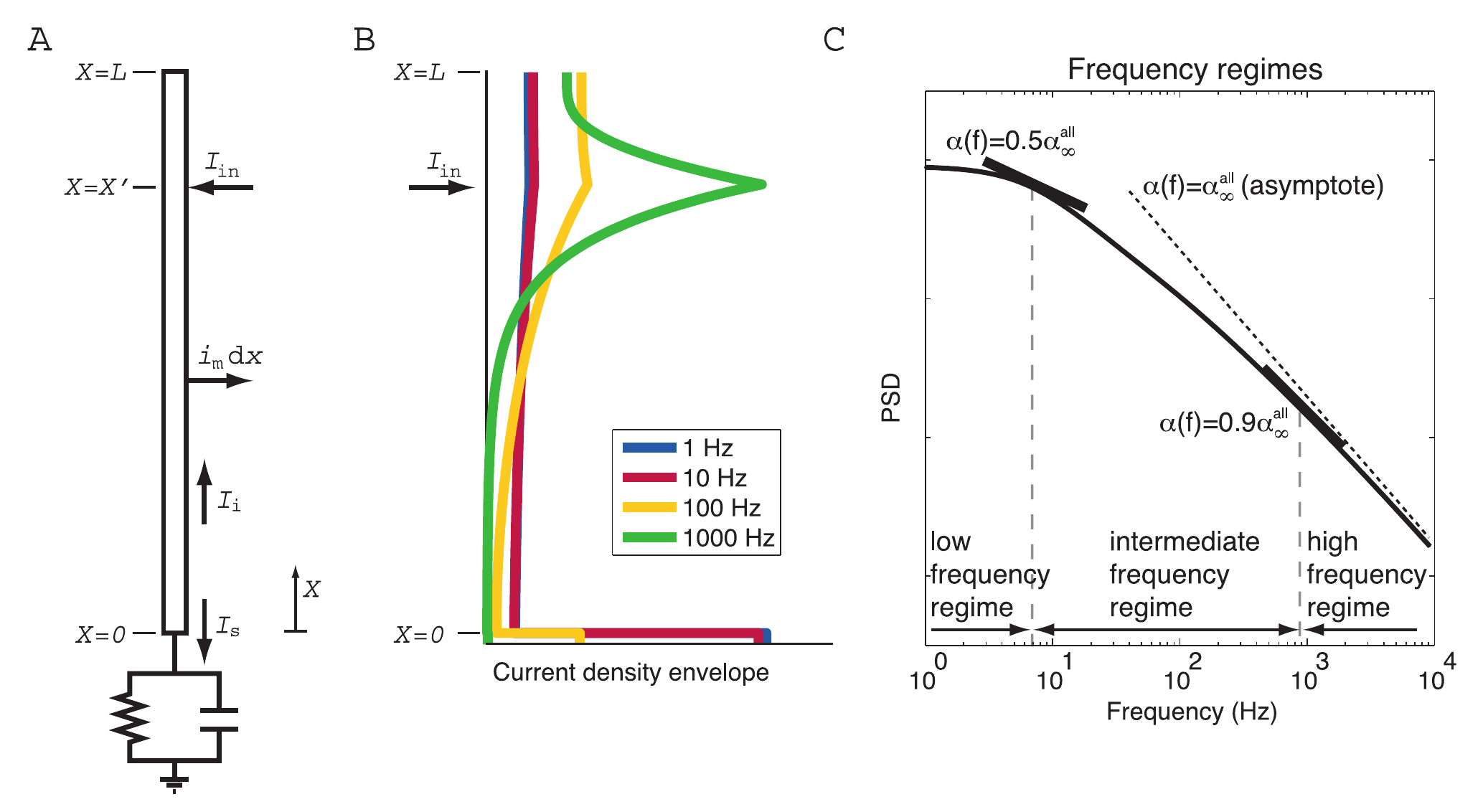}
\end{center}
\caption{
}
\label{fig:fig3X}
\end{figure}

\noindent
{\bf Figure~\ref{fig:fig3X}:} Schematic illustration of the ball and stick neuron model and its filtering properties.  (A) Schematic illustration of the ball and stick neuron model with a single input at a given position $X=X'$. The lumped soma is assumed iso-potential and located at $X=0$.
(B) Frequency-dependent current-density envelopes of return currents for a ball and stick neuron with input at $X=0.8L$. 
The somatic return currents are illustrated as current densities from a soma section with length $20~\mathrm{\mu m}$ placed below the stick. For 1~Hz, 10~Hz, 100~Hz and 1000~Hz the amplitudes of the somatic return currents are about 1/7.3, 1/7.5, 1/22 and 1/3100 of the input current, respectively. Parameters used for the ball and stick neuron model: stick diameter $d=2~\mathrm{\mu m}$, somatic diameter $d_\s=20~\mathrm{\mu m}$, stick length  $l=1~$mm, specific membrane resistance $R_\m=3~\Omega \mathrm{m}^2$, inner resistivity  $R_\ii=1.5~\Omega$m and specific membrane capacitance of $C_\m=0.01~\mathrm{F/m^2}$. This parameter set is the default parameter set used in the present study, see Table~\ref{tab:glossary}. (C) Representative log-log plot for a PSD when input is homogeneously distributed across the entire neuron model. The low frequency (\emph{lf}), intermediate frequency (\emph{if}) and high frequency (\emph{hf}) regimes are stipulated. 
The regimes are defined relatively to $\alpha^\mathrm{all}_\infty$ describing the asymptotic value of the exponent of the respective power-law transfer functions ($H^I$, $H^p$ or $H^V$),
with both uncorrelated and correlated input ('all' types of input) onto both the soma and the stick.

\

\begin{figure}[!ht]
\begin{center}
\includegraphics[width=6in]{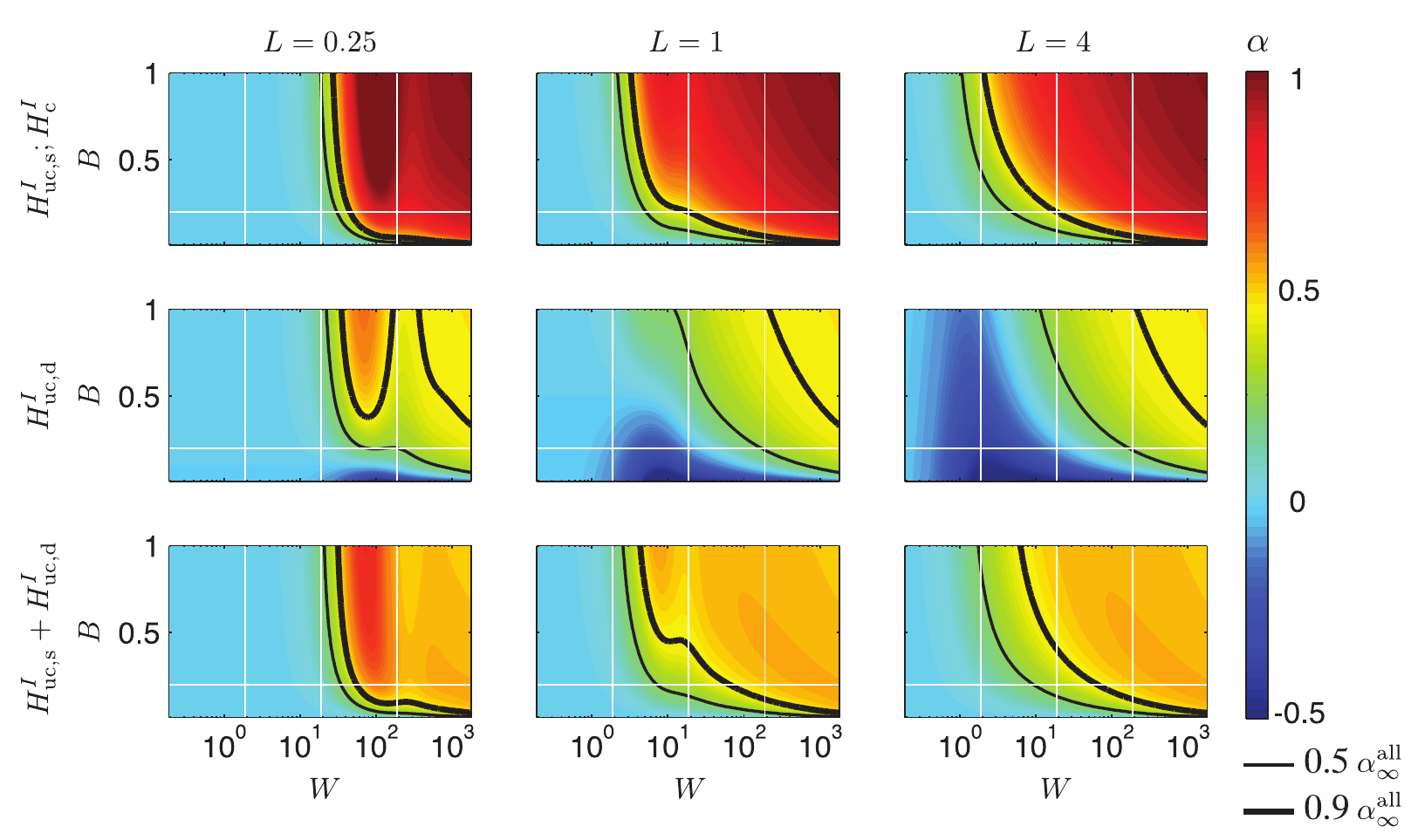}
\end{center}
\caption{
}
\label{fig:fig4X}
\end{figure}

\noindent
{\bf Figure~\ref{fig:fig4X}:} 
Slopes $\alpha^I$ for the PSD transfer function for the soma current for a ball and stick neuron in terms of its dimensionless parameters.
Row~1 corresponds both to correlated input currents ($H_\cc^I$) with any input densities $\rho_\s \ne \rho_\dd$, and to uncorrelated input to soma only ($H_{\uc,\s}^I$). Row~2 corresponds to the case of uncorrelated input currents solely onto the dendrite.  Row~3 corresponds to uncorrelated input currents with equal density, $\rho_\s=\rho_\dd$, throughout the neuron. The dimensionless parameter $B=d_\s^2/d \lambda$ is plotted along the vertical axes, while the dimensionless frequency $W$ is plotted logarithmically along the horizontal axes. In the left column the dimensionless length is $L=0.25$, in the middle column $L=1$ and the right column $L=4$. The horizontal white line express the default value of the parameter $B$, $B=0.2$
(soma diameter $d_\s=20~\mathrm{\mu m}$, stick diameter $d=2~\mathrm{\mu m}$, length constant $\lambda=1$~mm),
while the vertical white lines correspond to frequencies of 10~Hz, 100~Hz and 1000~Hz, respectively, for the default membrane time constant $\tau_\m=30$~ms. The thin black line denotes $\alpha=0.5 \alpha^\mathrm{all}_\infty=0.25$ and the thicker black line denotes $\alpha=0.9 \alpha^\mathrm{all}_\infty=0.45$, with $\alpha^\mathrm{all}_\infty=0.5$ denoting the asymptotic value for the case of both uncorrelated and correlated input onto the whole neuron. All plots use the same color scale for $\alpha$, given by the color bar to the right.

\

\begin{figure}[!ht]
\begin{center}
\includegraphics[width=6in]{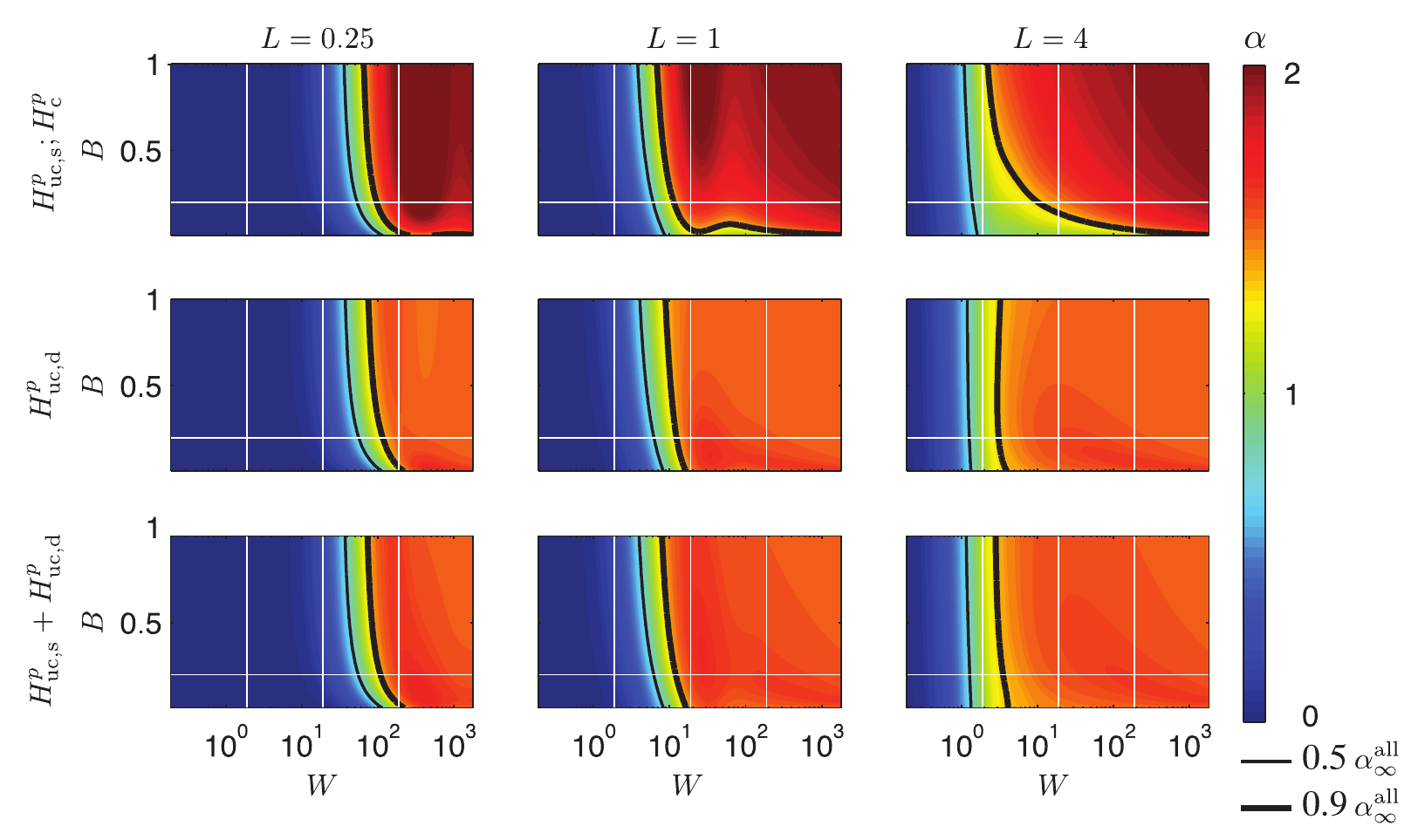}
\end{center}
\caption{
}
\label{fig:fig5X}
\end{figure}

\noindent
{\bf Figure~\ref{fig:fig5X}:} 
Slopes $\alpha^p$ for the PSD transfer function for the current-dipole moment (single-neuron EEG contribution) for a ball and stick neuron in terms of its dimensionless parameters.
Row~1 corresponds both to correlated input currents ($H_\cc^p$) with any input densities $\rho_\s \ne \rho_\dd$, and to uncorrelated input to soma only ($H_{\uc,\s}^p$). Row~2 corresponds to the case of input currents solely onto the dendrite. Row~3 corresponds to uncorrelated white-noise input currents with equal density, $\rho_\s=\rho_\dd$, throughout the neuron. The dimensionless parameter $B$ is plotted along the vertical axes, while the dimensionless frequency $W$ is plotted logarithmically along the horizontal axes. In the left column the dimensionless length is $L=0.25$, in the middle column $L=1$ and the right column $L=4$. The horizontal white line express the default value of the parameter $B$, $B=0.2$
(soma diameter $d_\s=20~\mathrm{\mu m}$, stick diameter $d=2~\mathrm{\mu m}$, length constant $\lambda=1$~mm), while the vertical white lines correspond to frequencies of 10~Hz, 100~Hz and 1000~Hz for the default membrane time constant $\tau_\m=30$~ms.
The thin black line denotes $\alpha=0.5 \alpha^\mathrm{all}_\infty=0.75$ and the thicker black line denotes $\alpha=0.9 \alpha^\mathrm{all}_\infty=1.35$, with  $\alpha^\mathrm{all}_\infty=1.5$ denoting the asymptotic value for the case of both uncorrelated and correlated input onto the whole neuron. All plots use the same color scale for $\alpha$, given by the color bar to the right.

\

\begin{figure}[!ht]
\begin{center}
\includegraphics[width=6in]{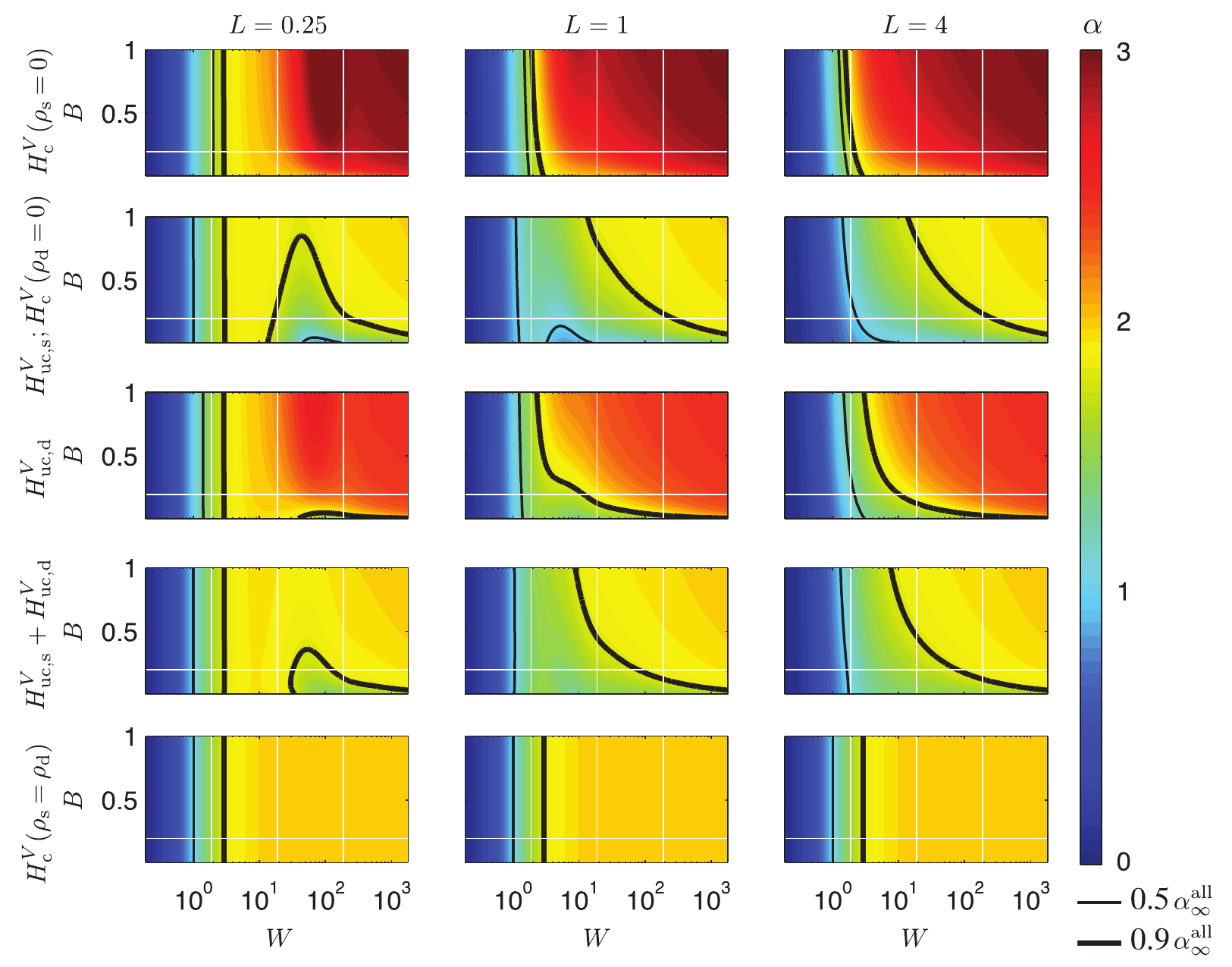}
\end{center}
\caption{
}
\label{fig:fig6X}
\end{figure}

\noindent
{\bf Figure~\ref{fig:fig6X}:} 
Slopes $\alpha^V$ for the PSD transfer function for the soma potential for a ball and stick neuron in terms of its dimensionless parameters.
Row~1 corresponds to correlated input currents solely onto the dendrite.  Row~2 corresponds to input currents solely onto soma, either correlated ($H^V_\cc(\rho_\dd=0)$) or uncorrelated ($H^V_{\uc,\s}$). In row~3 uncorrelated input currents are applied homogeneously across the dendrite. Row~5 corresponds to uncorrelated input currents with equal density, $\rho_\s=\rho_\dd$, throughout the neuron. Row~6 shows results for correlated input currents with equal density, $\rho_\s=\rho_\dd$, throughout the neuron. The dimensionless parameter $B$ is plotted along the vertical axes, while the dimensionless frequency $W$ is plotted logarithmically along the horizontal axes. In the left column the dimensionless length is $L=0.25$, in the middle column $L=1$ and the right column $L=4$. The horizontal white line express the default value of the parameter $B$, $B=0.2$
(soma diameter $d_\s=20~\mathrm{\mu m}$, stick diameter $d=2~\mathrm{\mu m}$, length constant $\lambda=1$~mm), while the vertical white lines correspond to frequencies of 10~Hz, 100~Hz and 1000~Hz for the default membrane time constant $\tau_\m=30$~ms. The thin black line denotes $\alpha=0.5 \alpha^\mathrm{all}_\infty=1$ and the thicker black line denotes $\alpha=0.9 \alpha^\mathrm{all}_\infty=1.8$, with  $\alpha^\mathrm{all}_\infty=2$ denoting the asymptotic value for the case of both uncorrelated and correlated input onto the whole neuron. All plots use the same color scale for $\alpha$, given by the color bar to the right.

\

\begin{figure}[!ht]
\begin{center}
\includegraphics[width=6in]{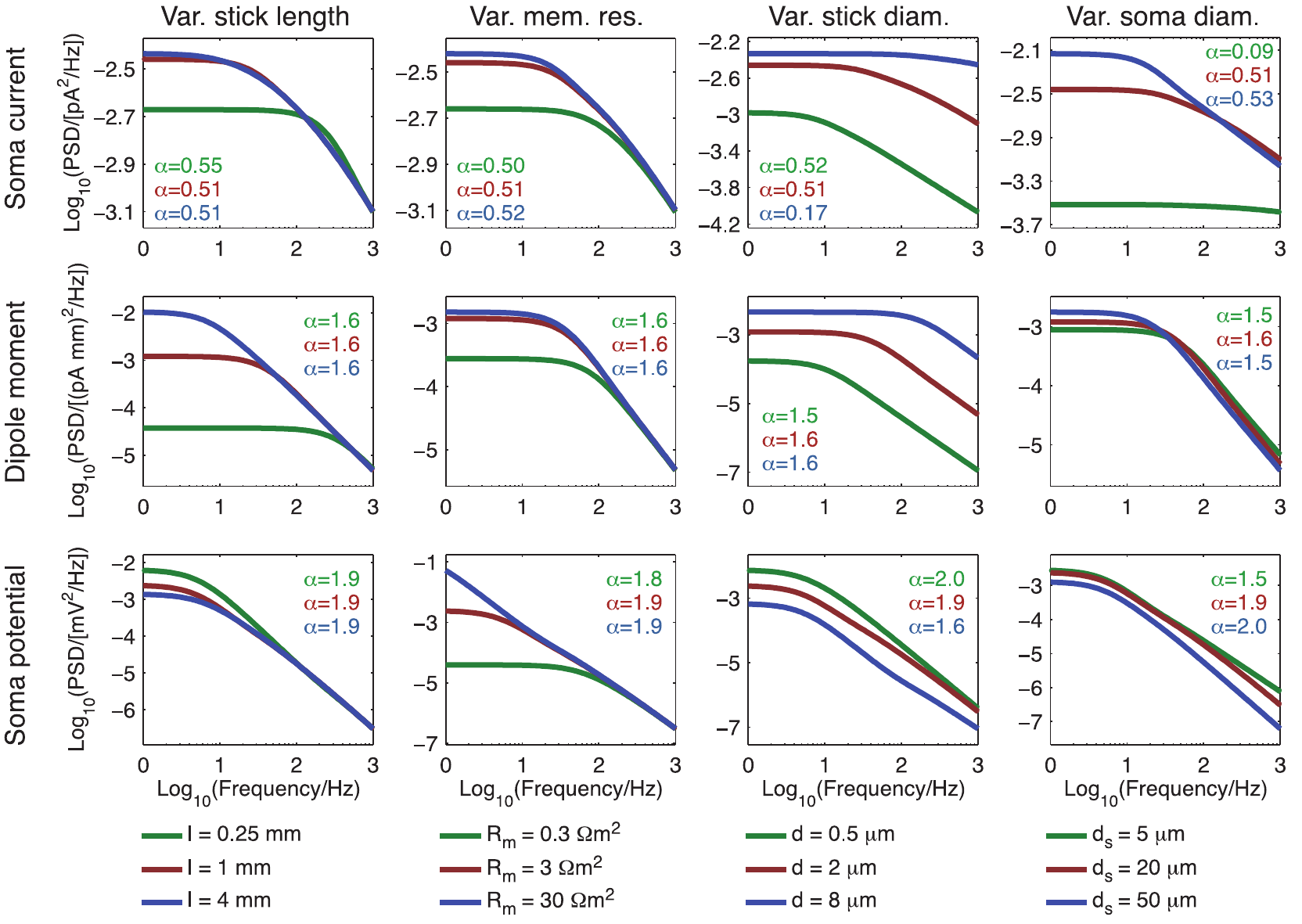}
\end{center}
\caption{
}
\label{fig:fig7X}
\end{figure}

\noindent
{\bf Figure~\ref{fig:fig7X}:} Dependence of PSDs on biophysical parameters for uncorrelated input. PSDs of the soma current (row 1), current-dipole moment (row 2) and soma potential (row 3)  for the ball and stick model with \emph{uncorrelated} white-noise input currents homogeneously distributed throughout the membrane. The input density is two inputs per square micrometer, and the input current is assumed to have a constant (white noise) PSD, $s= 1~\mathrm{fA^2/Hz}$. The columns show variation with stick length (first column), specific membrane resistance (second column),  stick diameter (third column) and soma diameter (fourth column) with values shown in the legends below the panels. All other parameters of the ball and stick neuron have default values: stick diameter $d=2~\mathrm{\mu m}$, somatic diameter $d_\s=20~\mathrm{\mu m}$, stick length $l=1~$mm, specific membrane resistance $R_\m=3~\Omega \mathrm{m}^2$, inner resistivity $R_\ii=1.5~\Omega$m and a specific membrane capacitance $C_\m=0.01~\mathrm{F/m^2}$. The values of $\alpha$ printed in the legends describe the powers of the slopes at 1000~Hz. The upper $\alpha$ corresponds to the low value of the parameter varied (green), the middle $\alpha$ corresponds to the default parameter (red), while the lower $\alpha$ corresponds to the high value of the parameter varied (blue).

\

\begin{figure}[!ht]
\begin{center}
\includegraphics[width=6in]{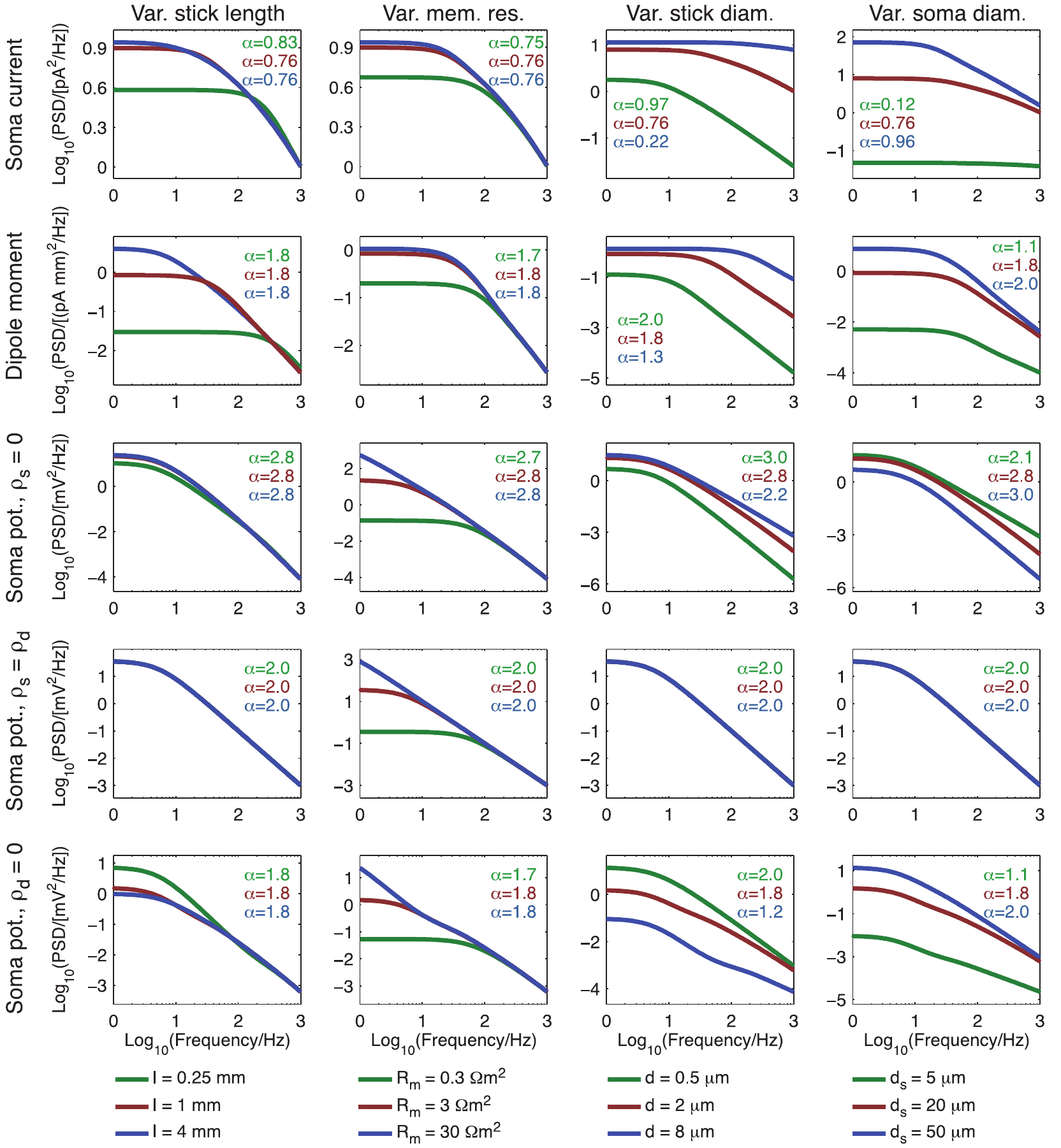}
\end{center}
\caption{
}
\label{fig:fig8X}
\end{figure}

\noindent
{\bf Figure~\ref{fig:fig8X}:} Dependence of PSDs on biophysical parameters for correlated input.  PSDs of the soma current (row 1), current-dipole moment (row 2) and soma potential (rows 3 to 5)  for the ball and stick model with \emph{correlated} white-noise input currents homogeneously distributed  throughout the stick only (row 1 to 3), the soma only (row 5) or with equal density throughout the soma and the stick (row 4). The input density is two inputs per square micrometer, unless a zero density is indicated on the axis. The input current is assumed to have a constant (white noise) PSD, $s= 1~\mathrm{fA^2/Hz}$. The columns show variation with stick length (first column), specific membrane resistance (second column),  stick diameter (third column) and soma diameter (fourth column) with values shown in the legends below the panels. All other parameters of the ball and stick neuron have default values: stick diameter $d=2~\mathrm{\mu m}$, somatic diameter $d_\s=20~\mathrm{\mu m}$, stick length $l=1~$mm, specific membrane resistance $R_\m=3~\Omega \mathrm{m}^2$, inner resistivity $R_\ii=1.5~\Omega$m and a specific membrane capacitance $C_\m=0.01~\mathrm{F/m^2}$. The values of $\alpha$ printed in the legends describe the powers of the slopes at 1000~Hz. The upper $\alpha$ corresponds to the low value of the parameter varied (green), the middle $\alpha$ corresponds to the default parameter (red), while the lower $\alpha$ corresponds to the high value of the parameter varied (blue).

\newpage

%

\clearpage
\newpage
\section{Tables}

\begin{table}[htdp]
\caption{List of symbols in alphabetical order. In the column labeled \emph{Default (Unit)} the default value of the parameter is given. If a default value is not listed,
the unit is given in parenthesis. The specific electrical properties of the soma membrane and stick membrane are here assumed to be equal.}
\begin{tabular}{lrl}
\hline
Symbol & Default (Unit) & Description \\
\hline
$B=d_\s^2/d \lambda$ & 0.2 & relative soma to infinite-stick conductance \\

$C_\m$ & $0.01~\mathrm{pF /\mu m^2}$ & specific membrane capacitance\\
$c_\m=\pi d C_\m$ & $0.0628~\mathrm{pF/\mu m}$  & membrane capacitance per unit length of cable\\
$d$ & $2~\mu\m$ & stick diameter\\
$d_\s$ & $20~\mu\m$ & soma diameter\\
$f$ & (Hz) & frequency \\
$G_\m=1/R_\m$  &  $0.333~\mathrm{pS /\mu m^2}$ & specific membrane conductance\\
$g_\m=1/r_\m=\pi d/R_\m$ & $2.09~\mathrm{pS/\mu m}$ & membrane conductivity per unit length of cable\\
$G_\infty=1/r_\ii \lambda$ & $2.09~\mathrm{nS}$ & infinite-stick conductance\\
$L=l/\lambda$ & 1 & electrotonic length\\
$l$ & $1~\m \m$ & stick length\\
$\comp{q}=\sqrt{1+jW}=\comp{Y}_\infty/G_\infty$ & (1) & frequency dependence of the infinite-stick admittance\\
$R_\ii$  & $1.5~\mathrm{M\Omega \mu m}$ & inner resistivity\\
$r_\ii=4 R_\ii/\pi d^2$ & $\mathrm{0.477~M\Omega/\mu m}$  & inner resistance per unit length of cable\\
$s$ & $1~\mathrm{fA^2/Hz}$ & power spectral density of input current\\
$T=t/\tau_\m$ & (1) & dimensionless time \\
$W=\omega \tau$ & (1) & dimensionless frequency \\
$X=x/\lambda$ & (1) & dimensionless position \\
$\comp{Y}_\mathrm{in}$ & ($\mathrm{S}$) & input admittance\\
$\comp{Y}=\comp{Y}_\s/\comp{Y}_\infty=\comp{q}B$ & (1) & relative soma to infinitestick admittance\\
$\comp{Y}_\s=\pi d_\s^2 G_\m \comp{q}^2$ & ($\mathrm{S}$) & soma admittance \\
$\comp{Y}_\infty=\comp{q}G_\infty$ & ($\mathrm{S}$) & infinite--stick admittance\\
$\lambda=1/\sqrt{g_\m r_\ii}$ &  $1~\m\m$ & neuron length constant\\
$\rho=\rho_\s/(\rho_\s+\rho_\dd)$  & 0.5 & relative input density\\
$\rho_\dd$ & $2/\mu \m^2$ & dendritic current-input number density \\
$\rho_\s$ &  $2/\mu \m^2$ & somatic current-input number density\\
$\tau_\m=R_\m C_\m$ & $30~\mathrm{ms}$ & membrane time constant\\
$\omega=2 \pi f$ & rad/s & angular frequency\\
\hline
\end{tabular}
\label{tab:glossary}
\end{table}%

\begin{table}[htdp]
\caption{PSD amplitudes and high-frequency power laws. The amplitudes $A$ and the asymptotic powers $\alpha_\infty$ for the different PSDs. The right column shows the amplitude $A'$ for the asymptotic PSDs expressed in terms of biophysical parameters. When $W$ approaches infinity, the asymptotic value of all PSD transfer functions except for $H_\cc^V$ is given by $H \rightarrow A W^{-\alpha_\infty}$. For $H_\cc^V$ there are two asymptotic values of non-standard form: $H_\cc^V \rightarrow A_\cc^V \rho^2 W^{-2}=\rho_\s^2 R_\m^2 W^{-2}$ for $\rho_\s \ne 0$ (left) and $H_\cc^V \rightarrow A_\cc^V B^{-2}W^{-3}=\rho_\dd^2 R_\m^2 B^{-2} W^{-3}$ for $\rho_\s=0$ (right). $(\ast)$: The values of the right column does not correspond to the given formula for $A'$, but rather to $A' \rho^2$ (left) and $A' B^{-2}$ (right).}
\begin{tabular}{lllc}
\hline
Case & Amplitude ($A$) & $\alpha_\infty$ ($W^{-\alpha_\infty}$) & $A'=A \times (f/W)^\alpha$\\
\hline
$H_\cc^I$ & $(\rho_\dd-\rho_\s)^2 (\pi d \lambda)^2$ & 1 & $(\rho_\dd-\rho_\s)^2  \pi d^3/8 R_\ii C_\m$\\
$H_{\uc,\dd}^I $ & $\rho_\dd \pi d  \lambda/\sqrt{2}$ & 1/2 & $\rho_\dd \pi^{1/2} d^{3/2}/4 R_\ii^{1/2}C_\m^{1/2}$\\
$H_{\uc,\s}^I$ & $\rho_\s \pi d \lambda/B$ & $1$  & $d^3 \rho_\s/8 C_\m d_\s^2 R_\ii$ \\
$H_\cc^p$ & $(\rho_\dd-\rho_\s)^2 \pi^2 d^2 \lambda^4$ & 2 & $(\rho_\dd-\rho_\s)^2 d^4/64 R_\ii^2 C_\m^2$\\
$H_{\uc,\dd}^p$ & $\rho_\dd  \pi d \lambda^3/\sqrt{2}$ & 3/2 & $\rho_\dd d^{5/2}/32\pi^{1/2} R_\ii^{3/2} C_\m^{3/2}$\\
$H_{\uc,\s}^p$ & $\rho_\s \pi d \lambda^3/B$ & $2$ & $d^4 \rho_\s/64 \pi  C_\m^2 d_\s^2 R_\ii^2$\\
$H_\cc^V$ & $(\rho_\dd+\rho_\s)^2 R_\m^2$ & 2;
3 & $\rho_\s^2/4 \pi^2C_\m^2$; $\rho_\dd^2 d^3/32 \pi^3 C_\m^3 d_\s^4 R_\ii~(\ast)$ \\
$H_{\uc,\dd}^V$ & $\rho_\dd R_\m^2/\sqrt{2} \pi d \lambda B^2$ & 5/2 & $\rho_\dd d^{3/2}/16 \pi^{7/2} d_s^4 R_\ii^{1/2} C_\m^{5/2}$\\
$H_{\uc,\s}^V$ & $\rho_\s R_\m^2/\pi d \lambda B$ & $2$ & $\rho_\s/4 \pi^3 C_\m^2 d_\s^2$ \\
\hline
\end{tabular}
\label{tab:series_parameters}
\end{table}%

\end{document}